\documentclass[11pt]{article}
\input epsf.sty
\usepackage{amsmath,amsthm,epsfig}
\usepackage{booktabs}
\usepackage{graphicx}
\usepackage{subfigure}
\usepackage{mathrsfs}
\usepackage{color}
\usepackage{bm}
\usepackage{float}
\usepackage{amsmath,amssymb}
\usepackage{natbib}
\usepackage{multirow}
\usepackage{tabularx}
\usepackage{calc}
\usepackage{geometry}
\usepackage{algpseudocode}
\usepackage{algorithm}
\RequirePackage{threeparttable} 

\newtheorem{proposition}{\bf{\sc Proposition}}

\newcommand{\ind}{\perp\!\!\!\!\perp} 
\emergencystretch=\maxdimen
\begin{document}
\title{\bf Model selection with uncertainty in estimating optimal dynamic treatment regimes}
\date{}
\author{Chunyu Wang \\
chunyu.wang@zju.edu.cn\\ Brian DM Tom \\
brian.tom@mrc-bsu.cam.ac.uk}
\maketitle

Optimal dynamic treatment regimes (DTRs), as a key part of precision medicine, have progressively gained more attention recently. To inform clinical decision making, interpretable and parsimonious models for contrast functions are preferred, raising concerns about undue misspecification.  It is therefore important to properly evaluate the performance of candidate interpretable models and select the one that best approximates the unknown contrast function. Moreover, since a DTR usually involves multiple decision points, an inaccurate approximation at a later decision point affects its estimation at an earlier decision point when a backward induction algorithm is applied.  This paper aims to perform model selection for contrast functions in the context of learning optimal DTRs from observed data. Note that the relative performance of candidate models may heavily depend on the sample size when, for example, the comparison is made between parametric and tree-based models. Therefore, instead of investigating the limiting behavior of each candidate model and developing methods to select asymptotically the `correct' one, we focus on the finite sample performance of each model and attempt to perform model selection under a given sample size. In other words, the estimand of interest is sample-size-specific.  To this end, we adopt the counterfactual cross-validation metric and propose a novel method to estimate the variance of the metric. Supplementing the cross-validation metric with its estimated variance allows us to characterize the uncertainty in model selection under a given sample size and facilitates hypothesis testing associated with a preferred model structure. Simulation studies are provided to demonstrate (i) the performance of our proposed variance estimator and (ii) the improvement achieved by incorporating model selection for contrast functions in estimating optimal DTRs. We apply our method to the analysis of Sequential Treatment Alternatives to Relieve Depression (STAR*D) data.

\section{Introduction}
By recognizing the heterogeneity in treatment effects (HTE) across individuals and seeking to adapt to the evolving health status of an individual over time, a dynamic treatment regime (DTR) formalizes clinical decision making as a sequence of treatment rules each of which maps the  patient's  up-to-date characteristics to a recommended treatment action. Given a class of feasible DTRs, an optimal DTR is the one leading to the best benefit (defined by a user specified utility function \citep{orellana2010dynamic}), on average, if followed by the population of concern. Pioneered by  \cite{murphy2003optimal} and \cite{robins2004optimal}, there is now an abundance of methodological research on learning optimal DTRs from observed data \citep{li2023optimal,kosorok2019precision,wang2025tutorial}. Generally, methods for learning optimal DTRs can be classified into indirect regression-based methods and direct-maximization (also called policy-search or value-search) methods \citep{laber2014dynamic,laber2015tree}. An indirect method infers optimal DTRs by modeling some aspects of the conditional distribution of the outcome, for example, conditional mean functions in Q-learning \citep{schulte2014q}, contrast functions or regret functions in A-learning \citep{moodie2007demystifying}; or modeling the marginal relationship between the outcome and candidate DTRs, such as the dynamic regime marginal structural models (drMSM) in \cite{robins2008estimationmsm} and \cite{orellana2010dynamic}. Such indirect methods heavily rely on the specification of the (conditional or marginal) regression models  and the optimal DTRs are then derived from the estimated regression models. 

Direct methods work on the regime-related maximization problem for which the maximizer over a prespecified class is the estimated optimal DTR. Examples include the (augmented) inverse probability weighted estimator ((A)IPWE) developed in \cite{zhang2013robust}, backward OWL (BOWL) and simultaneous OWL (SOWL) in \cite{zhao2015new}, augmented outcome-weighted learning (AOL) in \cite{liu2018augmented} and tree-based reinforcement learning (T-RL) in \cite{tao2018tree}. Methods such as OWL and AOL transfer the regime-related
maximization problem to a weighted classification problem, and hence these direct methods are also referred to as classification-based methods \citep{zhang2015using}. Depending on the algorithm utilized in learning optimal DTRs, the aforementioned methods are implemented either simultaneously or sequentially via backward induction. See Table \ref{tab:summary} for a summary of these standard methods. In general, the backward induction procedure is more computationally efficient than the simultaneous approach, since a single separate treatment rule is required to be estimated in each stage. However, model misspecification needs to be managed carefully while performing backward
induction, as an inaccurate estimation at a later decision point will then affect its estimation at an earlier decision point.  Direct methods, compared to indirect methods, rely on fewer model assumptions and are thus more robust to model misspecification \citep{zhang2015using}. From this point of view, direct methods are better choices for estimating optimal DTRs in practice. 

However, there are cases where clinical scientists wish to determine why the estimated regime is the optimal one and which subgroup of patients may benefit the most from following the estimated regime. This information is essential in informing future clinical trials for validation and in promoting the application of data-driven treatment regimes in clinical practice. Indirect methods usually provide insights into the aforementioned questions. Here we focus on the indirect and backward induction methods and attempt to incorporate model selection to mitigate against undue model misspecification, which is of main concern when applying such methods in practice. Note that A-learning, compared to Q-learning, is usually less prone to model misspecification as only the treatment-covariate interactions, expressed in terms of contrast functions or regret functions, are required to be specified. We follow the line of research on contrast functions and use cross-validation to perform model selection for contrast functions. 
\begin{table}[H]
\centering
\caption{\label{tab:summary}Summary of standard methods for estimating optimal DTRs.} \vspace{1mm}
\begin{tabular}{c|c|c}
\toprule 
\multicolumn{1}{c}{}&\multicolumn{1}{c}{simultaneous}  & \multicolumn{1}{c}{backward induction} \\ \hline 
direct & SOWL, (A)IPWE&BOWL, AOL, T-RL \\ \hline 
indirect &drMSM  &Q-learning, A-learning\\
\hline
\end{tabular}
\end{table}

The contrast function in DTRs is also called the conditional average treatment effect (CATE) in a single stage treatment regime. Several methods have been developed for model evaluation and selection on CATE, among which cross-validation and its variants are most straightforward and flexible tools \citep{schuler2018comparison, saito2020counterfactual,rolling2014model,claeskens2003focused}. Unlike cross-validation in the conventional regression context, the label for CATE is not readily available, and a pseudo-label is constructed in order to apply the cross-validation procedure for CATE. See \cite{saito2020counterfactual} and \cite{schuler2018comparison} for investigations on constructing the pseudo-label and discussions on existing counterfactual cross-validation metrics. Most literature on cross-validation (standard or counterfactual) for model selection focuses on its asymptotic behavior -- whether and when the proposed cross-validation method selects the asymptotically better model (or asymptotically best if there are more than two candidate models) with probability approaching one. Such theoretical investigations concern the scenario in which the true model is included in the set of candidate models \citep{shao1993linear} or the asymptotically better/best model is well-defined \citep{yang2007consistency,rolling2014model}. However, there are cases where each candidate model is an approximation to the true function, which are usually encountered when interpretation rather than prediction has been emphasized in application. Moreover, when the statistical properties of the candidate models differ markedly, for example, one parametric and pre-specified while the other nonparametric and adaptive, their relative performance in approximating the true function heavily depends on the sample size. From a practical point of view, it is desirable to select the better one under the given sample size. 

\begin{figure}[H]
    \centering
       \subfigure[]{\includegraphics[width=0.32\textwidth]{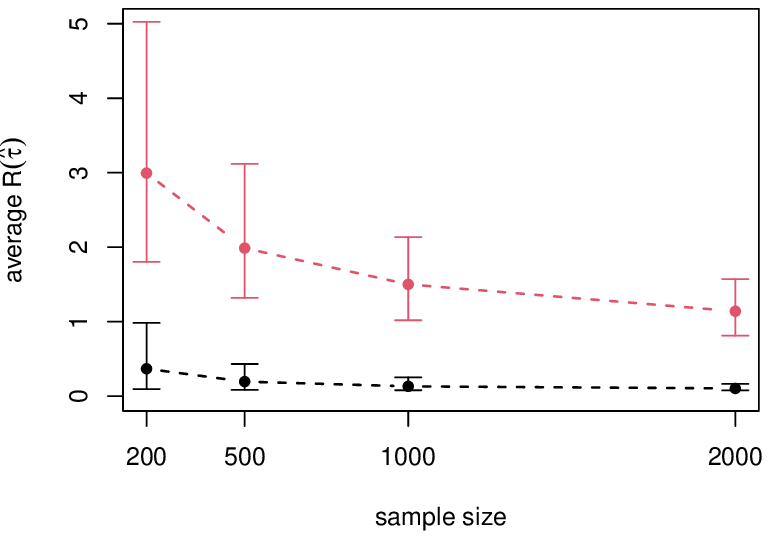}} 
    \subfigure[]{\includegraphics[width=0.32\textwidth]{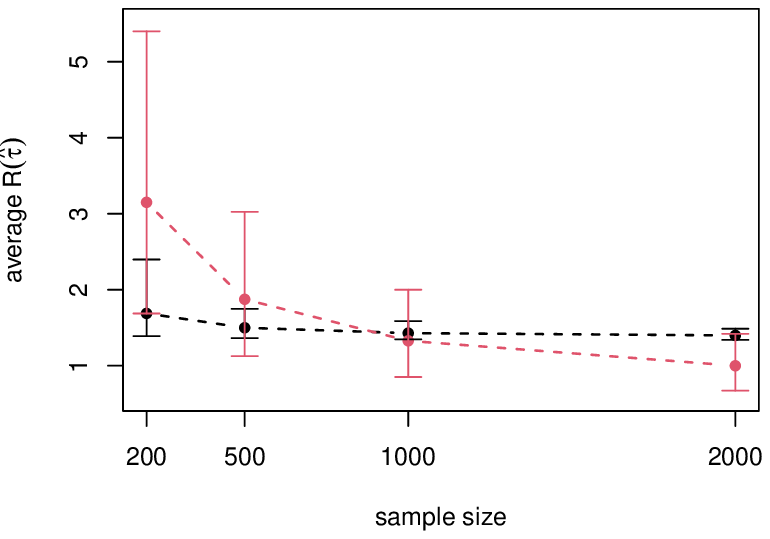}} 
    \subfigure[]{\includegraphics[width=0.32\textwidth]{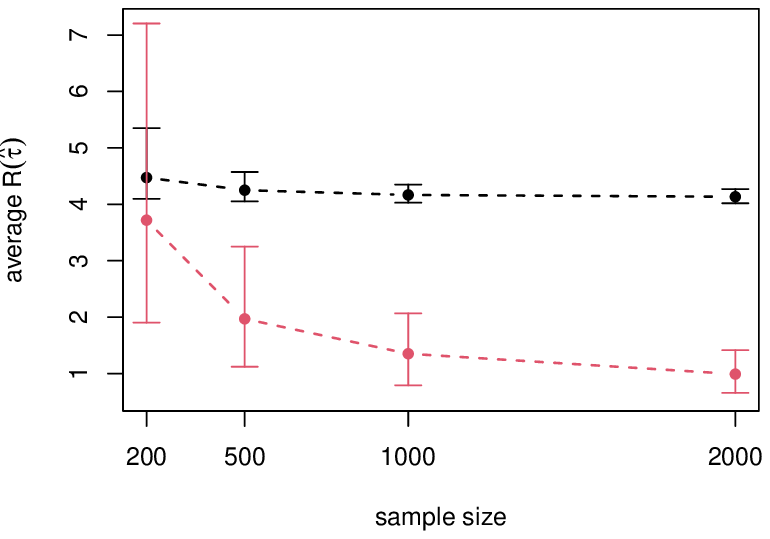}} \\
     \subfigure[$\tau(l)$ in (a)]{\includegraphics[width=0.32\textwidth]{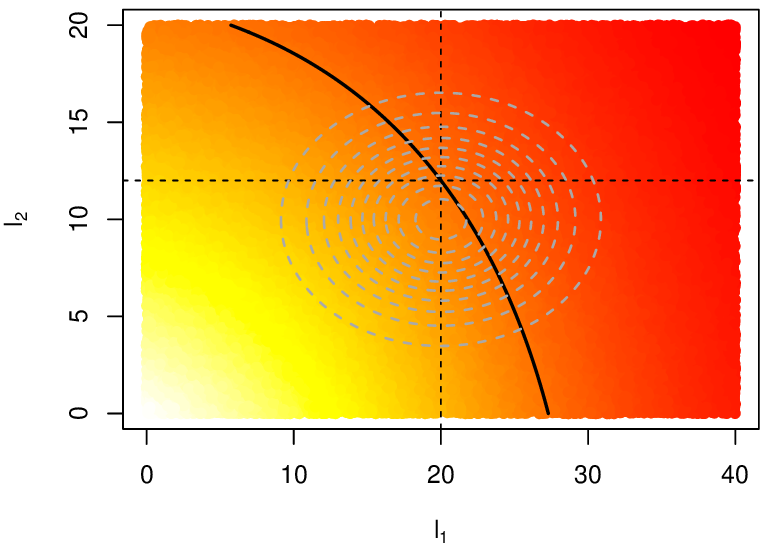}} 
     \subfigure[$\tau(l)$ in (b)]{\includegraphics[width=0.32\textwidth]{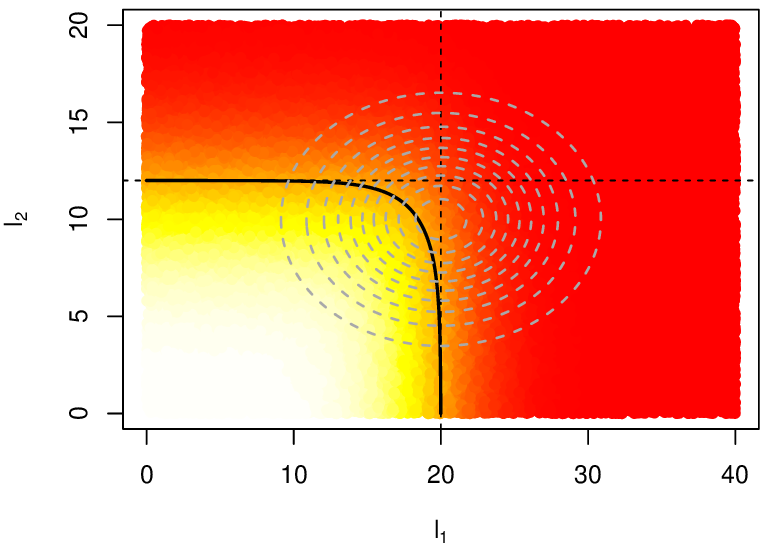}} 
       \subfigure[$\tau(l)$ in (c)]{\includegraphics[width=0.32\textwidth]{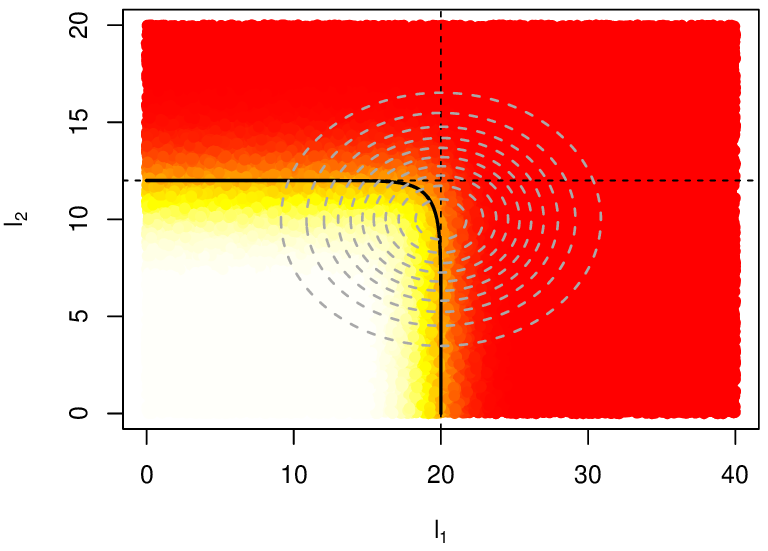}}
    \caption{\label{fig:mse_samplesize} The average $R(\hat{\tau})$ of a linear model (black) and a tree-based model (red) under different sample sizes and different true $\tau$'s. In subfigures (a)-(c), the average $R(\hat{\tau})$'s  (solid points) are calculated under 1000 Monte Carlo replications and the error bars indicate the 0.975 and 0.025 quantiles among those 1000 Monte Carlo replications. Subfigures (d)-(f) are the profiles of the true $\tau(l)$'s behind figures (a)-(c), respectively. Higher values of $\tau(l)$ are represented in deeper red  in (d)-(f) and the black solid curves  correspond to the boundaries $\{l: \tau(l)=0\}$. Grey dashed curves in (d)-(f) are contour plots of the density function of $(L_1,L_2)$.}
    \end{figure}

Consider the simulated scenario illustrated in Figure \ref{fig:mse_samplesize}. The subfigures (d)-(f) in Figure \ref{fig:mse_samplesize} give the true CATE profiles with respect to two biomarkers $(L_1,L_2)$, under three different settings. Two candidate models are considered under each setting: one is the linear model and the other is the tree-based model (considering a single tree and not a forest for interpretability), with their performances shown in black and red, respectively, in subfigures (a)-(c) under different sample sizes. In particular, an intersection occurs around $n=1000$ in subfigure (b) when the profile of CATE, as shown in subfigure (e), is neutral between the linear and tree structures. That is, the conclusion on model selection in this setting is sample-size-specific. Our objective, to re-iterate, is to determine which model performs better for CATE under a given sample size.

When making a model selection decision under a given sample size based on a counterfactual cross-validation metric, it is important to account for the uncertainty of performing finite-sample inference. Basically, the uncertainty of a cross-validation metric is two-fold \citep{lei2020cross}. The first is the uncertainty in the estimator (or the fitted model) learned from the training set $D_{tr}$, including the randomness due to sample splitting and the randomness of the observed data $D$, which is a random sample of fixed size drawn from the underlying distribution. While the randomness due to sample splitting can be well controlled by averaging over multiple splits, the randomness of the observed data cannot be tackled in such a way. The vertical lines (black and red error bars) in subfigures~\ref{fig:mse_samplesize} (a)-(c) reflect the variation of the estimators generated from 1000 Monte Carlo replications. 

The second part of the uncertainty comes from the label or pseudo-label in the validation set, leading to the overfitting issue with cross-validation when the decision is made based on the metric alone \citep{shao1993linear,lei2020cross}. To quantify this uncertainty, it is natural to estimate the variance of the cross-validation metric. However, unlike the cross-validation metric itself, estimating its variance is not straightforward. There are limited studies on this topic, all of which are developed within the conventional regression context \citep{NIPS1999,bengio2003no,markatou2005analysis,wang2014variance}. \cite{markatou2005analysis} proposed the method of moment approximation for the variance estimation. Their method depends on the specific form of the estimator being learned from the training set. Moreover, the incorporation of pseudo-labels in CATE complicates the moment approximation. \cite{wang2014variance} developed an unbiased variance estimator for a general U-statistic, which can be applied to a special form of the cross-validation metric under the restriction that the size of the training set is less than half the size of the original set. As the conclusion on model selection may differ when the sample size gets halved, this restriction is not applicable to our interested setting. 

Inspired by the variance decomposition formula for the average of exchangeable variables, we propose to estimate the variance of the counterfactual cross-validation metric by calculating the common correlation among different sample splittings via a data-adaptive approach. Supplementing the cross-validation metric with its variance enables finite-sample inference for model selection for CATE and similarly for the contrast functions in DTR. 

The remainder of this paper is organized as follows. Section 2 formulates the counterfactual Monte Carlo cross-validation metric and develops the variance estimator by targeting the common correlation. A half-and-half sample splitting method is proposed to approximate the correlation, and an inflation factor is considered to adjust the approximation. Section 3 extends the result for CATE to the DTR setting for contrast functions and develops the model selection-assisted A-learning for estimating optimal DTRs. Section 4 presents results from simulation studies to demonstrate the performance of the
proposed variance estimator and the improvement achieved by incorporating model
selection for contrast functions when estimating optimal DTRs. Finally, we apply our
method to the analysis of the `Sequential Treatment Alternatives to Relieve Depression'
(STAR*D) trial in Section 5.

\section{Model selection for a single decision point} \label{sec:model_selection_single}
Suppose we have observed data $D=\{(L_i,A_i,Y_i)\}_{i=1}^n$ collected from $n$ individuals in a randomized trial or an observational study. For the $i$th individual,  $L_i$ is a vector of pre-treatment covariates/biomarkers, $A_i\in\{0,1\}$ is the treatment indicator and $Y_i$ is the observed outcome for which, without loss of generality, larger values indicate better rewards. We assume that $(L_i,A_i,Y_i), i=1,\dots,n$ are independent and identically distributed. We drop the subscript $i$ to denote the random variable itself and use the corresponding lower-case letter to denote the value/realization of the random variable. Let $Y^{\ast}(a)$ denote the potential outcome that would have been observed if the individual had been assigned to treatment $a$, for $a=0,1$. Define $\tau(l)$ to be the conditional expectation of $Y^{\ast}(1)-Y^{\ast}(0)$ given $L=l$, i.e.,
$$\tau(l)=\mathbb{E}[Y^{\ast}(1)-Y^{\ast}(0)|L=l],$$ which is also called the conditional average treatment effect (CATE) or treatment effect function. Throughout this section, we make the following assumptions to ensure the identification of $\tau(l)$ from the observed data: (i) $Y^{\ast}(a)=Y$ if $A=a$, that is, the observed outcome is exactly the potential outcome under the treatment actually received, which is called the consistency assumption; (ii) the no unmeasured confounders assumption, i.e., $\{Y^{\ast}(0),Y^{\ast}(1)\}\ind A |L$; and (iii) the stable unit treatment value assumption, that is, the potential outcomes of any individual are unaffected by the treatment assignment of any other individual.

\subsection{From $\tau$-risk to cross-validation}
We aim to do model/estimator selection for $\tau:=\tau(\cdot)$ and to identify the best estimator $\hat{\tau}^{\star}$ amongst the $M$ candidate estimators $\hat{\tau}_{1},\dots, \hat{\tau}_{M}$. To this end, we begin by introducing some metrics for comparing estimators. For a given estimator $\hat{\tau}$ which is learned from the data at hand, the mean squared error   $R(\hat{\tau})=\mathbb{E}[\{\hat{\tau}(L)-\tau(L)\}^2|\hat{\tau}]$ is a natural metric to quantify the performance of $\hat{\tau}$ in approximating $\tau$, which is also referred to as the $\tau$-risk \citep{schuler2018comparison,doutreligne2023select}. The expectation involved in $R(\hat{\tau})$ is taken with respect to the distribution of $L$, which is independent of $\hat{\tau}$. The average of $R(\hat{\tau})$, i.e., $\mathbb{E}[R(\hat{\tau})]$, characterizes the average performance of $\hat{\tau}$ over hypothetical data sets drawn from the same distribution. In other words, $\mathbb{E}[R(\hat{\tau})]$ corresponds to the general performance (i.e. repeatability) of the algorithm behind $\hat{\tau}$ rather than its specific performance on the data at hand. Note that both $R(\hat{\tau})$ and $\mathbb{E}[R(\hat{\tau})]$ are unusable (in practice) since they depend on the unknown true $\tau$. We therefore need a surrogate for $\tau$, which in the conventional regression setting where the conditional mean function of $Y$ is of interest, is directly chosen to be $Y$ itself. 

For a specific surrogate $\widetilde{Y}$, it is straightforward to obtain the following relationship:
 \begin{equation}\label{surrogate}
    \mathbb{E}[\{\widetilde{Y}-\hat{\tau}(L)\}^2|\hat{\tau}]=\mathbb{E}[\{\widetilde{Y}-\tau(L)\}^2]+R(\hat{\tau})+2\mathbb{E}[\{\widetilde{Y}-\tau(L)\}\{\tau(L)-\hat{\tau}(L)\}|\hat{\tau}],
\end{equation}
where the expectation is taken with respect to the distribution of $(\widetilde{Y},L)$, which does not depend on $\hat{\tau}$ or equivalently does not depend on the data used to get $\hat{\tau}$. A reasonable surrogate $\widetilde{Y}$ should satisfy $\mathbb{E}[\widetilde{Y}-\tau(L)|L]=0$ so that the third term on the RHS of (\ref{surrogate}) is zero and thus $\mathbb{E}[\{\widetilde{Y}-\hat{\tau}(L)\}^2|\hat{\tau}]$ equals $R(\hat{\tau})$ up to an additive constant (irrelevant to $\hat{\tau}$). In fact, $\widetilde{Y}$ itself is a function of $(L,A,Y)$, that is, $\widetilde{Y}:=\widetilde{Y}(L,A,Y)$. There are several proposed constructions of $\widetilde{Y}$, for example, the transformed outcome $\widetilde{Y}=AY/\pi(L)-(1-A)Y/(1-\pi(L))$ with $\pi(L)=P(A=1|L)$. See \cite{schuler2018comparison} and \cite{saito2020counterfactual} for related discussions on the choice of $\widetilde{Y}$. In this paper, we follow the idea in \cite{rolling2014model} and construct $\widetilde{Y}$ based on matched treated and control individuals. Specifically, $\widetilde{Y}=(2A-1)(Y-Y_{A^c})$, where $Y_{A^c}$ is the outcome of the most similar individual in the opposite treatment group to $A$.

Given a specific construction of $\widetilde{Y}$, it is straightforward to estimate $\mathbb{E}[\{\widetilde{Y}-\hat{\tau}(L)\}^2|\hat{\tau}]$ via sample splitting:
\begin{equation}\label{sample splitting}
\hat{R}=\frac{1}{\#I_{val}}\sum_{i \in I_{val}}\left[\widetilde{Y}_i-\hat{\tau}(L_i;D_{tr})\right]^2,
\end{equation}
where $D$ is split into a training set $D_{tr}$ and a validation set $D_{val}$, $I_{tr}$ and $I_{val}$ are the index
sets of $D_{tr}$ and $D_{val}$, respectively, with $I_{val}=I_{tr}^c$ (i.e. the complement of the set $I_{tr}$), and $\hat{\tau}$ is learned from the training set $D_{tr}$ and its performance is evaluated on the remaining set, i.e., the validation set $D_{val}$. Note that in the standard regression setting where $\widetilde{Y}_i$ is taken as $Y_i$, the terms in the sum in (\ref{sample splitting}) are conditionally independent given $D_{tr}$. This is not the case for CATE since the constructed $\widetilde{Y}_i$ for $\tau$ usually depends on observations additional to the $i$th observation $(L_i,A_i,Y_i)$. Consider the matching-based construction we adopt here, $\widetilde{Y}_i=(2A_i-1)(Y_i-Y_{i^{\prime}})$, with $i^{\prime}\in I _{val}$ being the most similar individual in the opposite treatment
group to individual $i$, which implies that individuals in the validation set are repeatedly used as the sum in (\ref{sample splitting}) goes over all individuals in $I_{val}$.  \cite{rolling2014model} proposed another scheme to maintain the conditional independence by firstly partitioning the space of $L$ into cells and then randomly selecting a pair of $(Y_i,Y_{i^{\prime}})$ within each cell. This method ensures that each individual in $I_{val}$ is used in at most one treatment-control pairing, which facilitates the investigation of the asymptotic properties of $\widehat{R}$. For ease of application, we stick with $\widehat{R}$ defined by (\ref{sample splitting}).

To reduce the variability incurred from sample splitting, Monte Carlo cross-validation 
(MCCV) repeatedly and randomly divides 
$D$ into training sets and validation sets, and averages the risk estimators $\widehat{R}$'s  calculated from different splits. Specifically, let $q$ be a prespecified proportion of individuals in the validation set and let $I_T$ and $I_C$ denote the index set of the treated and untreated/control individuals, i.e., $I_{T}=\{i: A_i=1,i=1,\dots,n\}$ and $I_{C}=\{i: A_i=0,i=1,\dots,n\}$. MCCV proceeds as follows: 
\begin{itemize}
    \item[(i)] Sample $q\cdot\#I_T$  and $q\cdot\#I_C$ individuals without replacement from $I_T$ and $I_C$ to get $I_{val,T}$ and $I_{val,C}$ respectively. Let $I_{val}=I_{val,T}\cup I_{val,C}$ and create the validation set $D_{val}=\{(L_i, A_i,Y_i), i\in I_{val}\}$ and the training set $D_{tr}=D\setminus D_{val}$;
    \item[(ii)] Train the candidate algorithms on $D_{tr}$ (or fit the candidate models to $D_{tr}$) to obtain $\hat{\tau}_{1}(\cdot;D_{tr}),\dots, \hat{\tau}_{M}(\cdot;D_{tr})$;
    \item[(iii)] Construct $\widetilde{Y}_i$ for $i \in I_{val}$ and calculate the risk estimator  $\widehat{R}=\sum_{i \in I_{val}}U_i(D_{tr})/\#I_{val}$ under the given sample splitting, where either $U_i(D_{tr})$ can be the squared-error loss, as in (\ref{sample splitting}), to assess the performance of a specific estimator, or $U_i(D_{tr})$ can be defined as the difference in performance between two estimators of interest, e.g., $U_i(D_{tr})=[\widetilde{Y}_i-\hat{\tau}_{lin}(L_i;D_{tr})]^2-[\widetilde{Y}_i-\hat{\tau}_{tre}(L_i;D_{tr})]^2$ when we are interested in the comparison of the linear model and the tree-based model/algorithm in estimating $\tau$; 
    \item[(iv)] Repeat steps (i)-(iii) 
$J$ times and average the 
$J$ risk estimators to obtain $\widehat{R}_{cv}=\sum_{j=1}^{J}\widehat{R}_{j}/J$, where $\widehat{R}_{j}$ is the risk estimator under the $j$th sample splitting.
\end{itemize}
  
\begin{figure}[H]
\centering
\includegraphics[height=3cm,width=12cm]{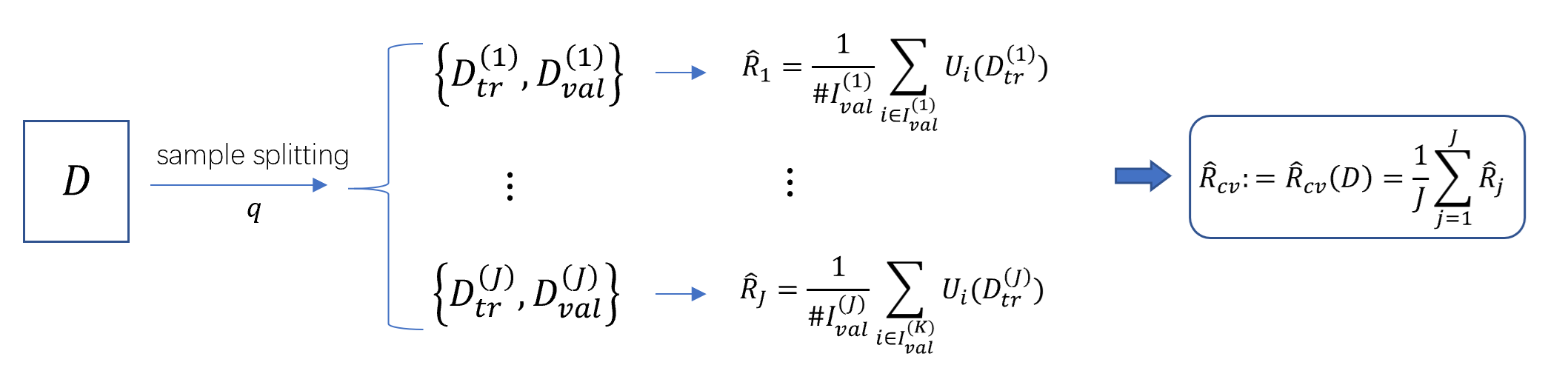}
\caption{\label{fig:mccv}Illustration of MCCV.}
\end{figure}

  See Figure \ref{fig:mccv} for an illustration of MCCV. Unlike the standard MCCV in which the splitting works on $D$ or equivalently on $\{1,\dots,n\}$, MCCV for CATE applies the splitting procedure to $I_T$ and $I_C$ separately and independently. In this way, the ratio of being treated to untreated in $D_{val}$ (also $D_{tr}$) remains the same as that in the original sample $D$, which facilitates the matching procedure between $D_{val,T}$ and $D_{val,C}.$ Note that $\widehat{R}_{cv}$, under valid construction of $\widetilde{Y}$, is an unbiased estimate of $\mathbb{E}[U_{n+1}(D_{n_{tr}})]$, with $n_{tr}=n(1-q)$; different from $\mathbb{E}[U_{n+1}(D)]$ due to the difference in sample size. (The subscript $n_{tr}$ indicates the sample size used to train the estimators/algorithms and the subscript $n+1$ is to emphasize the evaluation is performed on an independent observation.) As we are interested in the performance of candidate estimators under a given sample size $n$, it is reasonable to choose $n_{tr}$ close to $n$ so that $\widehat{R}_{cv}$ can reflect the corresponding performance under $n$. We here set $q=0.2$. Moreover, this inherent bias of $\widehat{R}_{cv}$ in estimating $\mathbb{E}[U_{n+1}(D)]$ also highlights the importance of taking the uncertainty of $\widehat{R}_{cv}$ into account. Consider the case shown in (b) in Figure \ref{fig:mse_samplesize} when $n=1000$. Choosing the linear model or the tree-based model in fitting $\tau$ in (e) does not make much difference and the decision making based on $\widehat{R}_{cv}$ involves a lot of uncertainty. When it comes to cases in (a) and (c) with $\tau$ given by (d) and (f) respectively, it is expected that $\widehat{R}_{cv}$ under $n=1000$ leads to selection with greater confidence. Therefore, it is important to supplement $\widehat{R}_{cv}$ with its variance to inform the uncertainty/confidence in the selection.

\subsection{Variance of $\widehat{R}_{cv}$}\label{sec:var_mccv}

$\widehat{R}_{cv}$ provides a straightforward point estimate of $\mathbb{E}[U_i(D_{n_{tr}})]$. However, an estimate of the variance of $\widehat{R}_{cv}$ is not straightforward to obtain. The complexity results from the dependence between $\widehat{R}_{1},\dots, \widehat{R}_{J}$, each of which is derived from a particular sample splitting of $D$.  Based on the fact that $\widehat{R}_{1},\dots, \widehat{R}_{J}$ are exchangeable  with a common positive correlation  $\text{corr}(\widehat{R}_{j},\widehat{R}_{j^{\prime}})=\rho,\forall j \neq j^{\prime}$,  \cite{NIPS1999} pointed out the following relationship between $\text{Var}[\widehat{R}_{cv}]$ and $\rho$:
\begin{equation}\label{variance_rho}
\text{Var}[\widehat{R}_{cv}]=\text{Var}[\widehat{R}_{1}]\left(\rho+\frac{1-\rho}{J}\right)=\text{Var}[\widehat{R}_{1}](1-\rho)\left(\frac{1}{J}+\frac{\rho}{1-\rho}\right),
\end{equation}
where $\text{Var}[\widehat{R}_{1}](1-\rho)$ can be estimated by $S_{R}^2$, the sample variance of $\{\widehat{R}_{1},\dots, \widehat{R}_{J}\}$ since
\begin{equation}\label{samplevariance_rho}
\mathbb{E}[S_{R}^2]=\text{Var}[\widehat{R}_{1}](1-\rho).
\end{equation}
It follows from (\ref{variance_rho}) and (\ref{samplevariance_rho}) that the variance of $\widehat{R}_{cv}$ can be estimated by $S_R^2(\frac{1}{J}+\frac{\hat{\rho}}{1-\hat{\rho}})$ if an unbiased estimate $\hat{\rho}$ of $\rho$ is provided. We therefore focus on developing estimators of $\rho$ subsequently. Before proceeding to $\hat{\rho}$ in the CATE setting, we firstly investigate the properties of $\rho$ when $\{U_i(D_{tr}^{(j)})\}_{i\in I_{val}^{(j)}}$ are conditionally independent given $D_{tr}^{(j)}$, which is exactly the case for cross-validation in the standard regression setting; and then clarify the extra complexities arising in the CATE setting. 

\begin{proposition} 
Let $\widehat{R}_{j}=\frac{1}{n_{2}} \sum_{i \in I_{val}^{(j)}}U_{i}(D_{tr}^{(j)})$  with $n_2=\#I_{val}^{(j)}=nq$ and assume that $\{U_{i}(D_{tr}^{(j)})\}_{i \in I_{val}^{(j)}}$ are conditionally independent given $D_{tr}^{(j)}$ for $j=1,\dots J$. Then we have the following expression for  $\rho=\text{corr}(\widehat{R}_{j},\widehat{R}_{j^{\prime}})$ with $j\neq j^{\prime}$:
\begin{equation}\label{rho}
\rho=\frac{\frac{1}{n}\rho_1+\frac{n-1}{n}\rho_3}
{\frac{1}{n_2}+\frac{n_2-1}{n_2}\rho_2},
\end{equation}
where $\rho_1=\text{corr}(U_i(D_{tr}^{(j)}),U_i(D_{tr}^{(j^{\prime})}))$ with $i \in I_{val}^{(j)}\cap I_{val}^{(j^{\prime})}$, $\rho_2=\text{corr}(U_i(D_{tr}^{(j)}),U_{i^{\prime}}(D_{tr}^{(j)}))$ with $i, i^{\prime}\in I_{val}^{(j)}$ and $i\neq i^{\prime}$, and $\rho_3=\text{corr}(U_i(D_{tr}^{(j)}),U_{i^{\prime}}(D_{tr}^{(j^{\prime})}))$ with $i \in I_{val}^{(j)}$, $i^{\prime} \in I_{val}^{(j^{\prime})}$ and $i\neq i^{\prime}$.
\end{proposition}
The proof is provided in Appendix A. We make the following remarks regarding the proposition:
\begin{itemize}
\item [1.]
  The correlation between $\widehat{R}_{j}$ and $\widehat{R}_{j^{\prime}}$ comes from two sources: (i) the correlation due to the training sets $D_{tr}^{(j)}$ and $D_{tr}^{(j^{\prime})}$; and (ii) the correlation due to evaluating on a common validation individual. In general, (ii) is considered to be the dominant source of the correlation. \cite{NIPS1999} proposed an approximation to $\rho$, $\hat{\rho}=n_2/n=q$, by taking  $U_i(D_{tr}^{(j)})$ as the function of the $i$th observation only, which is equivalent to setting $\rho_1=1$ and $\rho_2=\rho_3=0$ in (\ref{rho}). However, this approximation can overestimate or underestimate $\rho$, leading to a considerably biased estimator of $\text{Var}[\widehat{R}_{cv}]$. See Appendix B for a simulation to illustrate the issue.

 \item [2.]The decomposition given in (\ref{rho}) doesn't actually facilitate the estimation of $\rho$ in practice because each component, namely $\rho_1$, $\rho_2$ and $\rho_3$, is difficult to estimate with the combinations of $U_i(D_{tr}^{(j)}),i \in I_{val}^{(j)}, j=1,\dots, J$.
    
 \item [3.]   When considering $\rho$ in the CATE setting, $U_i(D_{tr}^{(j)})$ depends on $\widetilde{Y}_{i}$, which involves two individuals $i$ and $i^{\prime}$ in $I_{val}^{(j)}$ based on the pairing. Therefore, there are two possible cases for the correlation between $U_i(D_{tr}^{(j)})$ and $U_i(D_{tr}^{(j^{\prime})})$: (i) the individual $i \in I_{val}^{(j)}\cap I_{val}^{(j^{\prime})}$ matches with the same individual $i^{\prime}$ in the two different sample splittings $\{D_{tr}^{(j)},D_{val}^{(j)}\}$ and $\{D_{tr}^{(j^{\prime})},D_{val}^{(j^{\prime})}\}$; that is, $i^{\prime}$ being matched with $i$ in $I_{val}^{(j)}$ happens to be selected in $I_{val}^{(j^{\prime})}$ (with probability $q$) and also turns out to be the most similar individual to $i$ in the opposite treatment group in $I_{val}^{(j^{\prime})}$; (ii) the individual $i$ matches with two distinct individuals in $I_{val}^{(j)}$ and $I_{val}^{(j^{\prime})}$. It is therefore no longer reasonable to set $\rho_1=1$ in CATE. In fact, $q$ tends to overestimate $\rho$.
\end{itemize}

\begin{figure}[H]
\centering
\includegraphics[height=3.5cm,width=15cm]{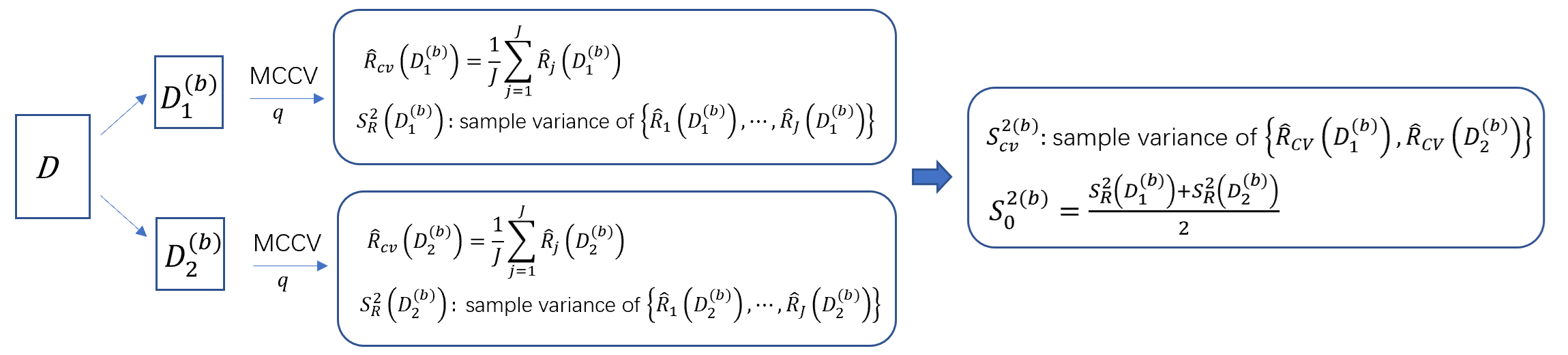}
\caption{\label{fig:mccv_rho} The $b$th half-and-half sample splitting.}
\end{figure}

 We propose to tackle the complexity of estimating $\rho$ in the CATE setting with a data-driven approach. Specifically, we randomly split  $D$ into two  equal sets  $D_1$ and $D_2$ by splitting $D_{T}$ ($D_C$) into two distinct data sets $D_{T,1}$ and $D_{T,2}$ ($D_{C,1}$ and $D_{C,2}$), each of size $\#I_{T}/2$ ($\#I_{C}/2$), and creating $D_1=D_{T,1}\cup D_{C,1}$ and $D_2=D_{T,2}\cup D_{C,2}$. Then we apply MCCV procedure with a fixed validation proportion $q$ to both $D_1$ and $D_2$, which leads to $\widehat{R}_{cv}(D_1)$ and $\widehat{R}_{cv}(D_2)$, as well as $S^{2}_{R}(D_1)$ and $S^{2}_{R}(D_2)$, the sample variances of $\{\widehat{R}_1(D_1),\dots, \widehat{R}_J(D_1)\}$ and $\{\widehat{R}_1(D_2),\dots, \widehat{R}_J(D_2)\}$ respectively. Notice that $\widehat{R}_{cv}(D_1)$ and $\widehat{R}_{cv}(D_2)$ are independent. The sample variance of $\widehat{R}_{cv}(D_1)$ and $\widehat{R}_{cv}(D_2)$, denoted by $S^2_{cv}$, therefore provides an unbiased (though noisy) estimate of $\text{Var}[\widehat{R}_{cv}^{n/2}]$, where the superscript $n/2$ is used to indicate the sample size of the whole set on which MCCV is conducted. Additionally, we let $S_0^{2}$ be the average of $S^{2}_{R}(D_1)$ and $S^{2}_{R}(D_2)$. See Figure \ref{fig:mccv_rho} for an illustration of the splitting procedure. We repeat the above steps $B$ times and average $\{S^{2(b)}_{cv}$: $b=1,\dots,B\}$ to get a good estimate of $\text{Var}[\widehat{R}_{cv}^{n/2}]$, which combined with the average from $\{S_0^{2(b)}$: $b=1,\dots, B\}$ leads to an estimate of $\rho$ derived from (\ref{variance_rho}) and (\ref{samplevariance_rho}) under the sample size $n/2$. That is 
\begin{equation}\label{rho_n/2}
    \hat{\rho}^{n/2}=1-\frac{1}{\bar{S}^2_{cv}/\bar{S}^2_{0}+1-1/J},
\end{equation}
where $\bar{S}^2_{cv}$ is the average of $S^{2(b)}_{cv}$'s for $b=1,\dots,B$, and $\bar{S}^2_{0}$ is the average of $S^{2(b)}_{0}$'s for $b=1,\dots,B$. Based on the assumption that the correlation obtained in the MCCV from the half sample size is close to that obtained in MCCV from the original sample size,  we propose to approximate $\rho$ with $\hat{\rho}^{n/2}$, which gives rise to an estimate of $\text{Var}[\widehat{R}_{cv}]$ as $S_R^2(\frac{1}{J}+\frac{\hat{\rho}^{n/2}}{1-\hat{\rho}^{n/2}})$, with $S_R^2$ being the sample variance of the original $\widehat{R}_{j}$'s for $j=1,\dots, J$. 

The rationale behind the aforementioned assumption lies in two points. Firstly, the correlation, compared to the variance itself, is more robust against the change of sample size when keeping the validation proportion $q$ unchanged. It is expected that approximating $\text{Var}[\widehat{R}_{cv}]$ directly with $\bar{S}^2_{cv}$ would yield considerable positive bias. \cite{NIPS1999} suggested another estimator of $\text{Var}[\widehat{R}_{cv}]$ based on $\bar{S}^2_{cv}$  using a similar half-and-half sample splitting procedure but performing MCCV in each splitting with the validation proportion $2q$. In this way, the size of validation set in the half-and-half sample splitting remains the same as that in the original set, i.e., $\#I_{val}(D_1)=\#I_{val}$. However, the size of the training set decreases substantially, leading to an overestimate of $\text{Var}[\widehat{R}_{cv}]$. Secondly, $\hat{\rho}^{n/2}$ is adaptive to the extra complexity in estimating $\rho$ in CATE, as  mentioned in remark 3, by using such a data-driven approach.

However, as indicated by the simulation results in Section \ref{sec:sim_var}, $\hat{\rho}^{n/2}$ tends to underestimate $\rho$ when $\hat{\tau}(\cdot,D_{tr})$ is sensitive to the perturbations in the training set, i.e., when $\hat{\tau}(\cdot,D_{tr}^{(j)})$'s vary a lot across $j$, which is the case for tree-based estimators. We therefore seek to adjust $\hat{\rho}^{n/2}$ with an inflation factor to accommodate the increased stability of $\hat{\tau}(\cdot,D_{tr})$ when the sample size gets doubled. Note that the expression of $\rho$ given by (\ref{rho}) can be rewritten as 
\begin{equation} \label{rho_again}
\rho=\frac{\frac{n_2}{n}\rho_1+\frac{n_2(n-1)}{n}\rho_3}
{1+(n_2-1)\rho_2},
\end{equation}
and that the increased stability of $\hat{\tau}(\cdot,D_{tr}^{(j)})$ (i.e., the decreased variation of $\hat{\tau}(\cdot,D_{tr}^{(j)})$) over $D_{tr}^{(j)}$ leads to a decreased value of $\rho_2=\text{corr}(U_i(D_{tr}^{(j)}),U_{i^{\prime}}(D_{tr}^{(j)}))$, which, if we assume the numerator in (\ref{rho_again}) remains almost the same, results in increased $\rho$. We therefore aim to characterize the change of $\rho_2$ when the sample size gets doubled and adapt $\hat{\rho}^{n/2}$  accordingly. 

Let $\rho^{n/2}$ and $\rho_2^{n/2}$ denote the counterparts of $\rho$ and $\rho_2$ in the half-sample case; that is,  $\rho^{n/2}=\text{corr}(\hat{R}_j(D_1),\hat{R}_{j^{\prime}}(D_1))$ and  $\rho_2^{n/2}=\text{corr}(U_i(D_{1,tr}^{(j)}),U_{i^{\prime}}(D_{1,tr}^{(j)}))$. Note that $\rho_2$ is reflected in the relationship between the variance of $\hat{R}_j$ and the variance of $U_i(D^{(j)}_{tr})$. Specifically, 
\begin{equation}\label{rho2}
    n_2\text{Var}[\hat{R}_1]/\text{Var}[U_i(D^{(1)}_{tr})]=1+(n_2-1)\rho_2, 
\end{equation}
 due to the equation $\text{Var}[\hat{R}_1]=\text{Var}[U_i(D^{(1)}_{tr})](\rho_2+(1-\rho_2)/n_2)$. Similarly,  for the half-and-half sample splitting case, we have 
 \begin{equation*}
 \frac{n_2}{2}\text{Var}[\hat{R}_1(D_1)]/\text{Var}[U_i(D^{(1)}_{1,tr})]=1+\left(\frac{n_2}{2}-1\right)\rho_2^{n/2},
 \end{equation*}
which together with (\ref{rho2}) leads to the following result
\begin{equation}\label{inflation_factor}
\frac{\text{Var}[\hat{R}_1(D_1)]/\text{Var}[U_i(D^{(1)}_{1,tr})]}{2\text{Var}[\hat{R}_1]/\text{Var}[U_i(D^{(1)}_{tr})]}=\frac{\text{Var}[\hat{R}_1(D_1)]}{2\text{Var}[\hat{R}_1]} \frac{\text{Var}[U_i(D^{(1)}_{tr})]}{\text{Var}[U_i(D^{(1)}_{1,tr})]}=\frac{1+(\frac{n_2}{2}-1)\rho_2^{n/2}}{1+(n_2-1)\rho_2}\approx \frac{\rho}{\rho^{n/2}},
\end{equation}
where the approximation in (\ref{inflation_factor}) comes from the earlier assumption that the numerator in (\ref{rho_again}) remains almost the same when the sample gets doubled. The product of ratios in (\ref{inflation_factor}) can thus be used as the inflation factor as it characterizes the change of $\rho$ due to the change of $\rho_2$ when the half-and-half sample splitting is applied. However, none of the four variances are known. We instead work on their corresponding sample variances. In addition to the sample variances $S_{R}^2$ and $\bar{S}_{0}^2$ introduced earlier, we define $S_{U}^2$ to be the average of the sample variances of $\{U_i(D_{tr}^{(j)})\}_{i \in I_{val}^{(j)}}$  over $j=1,\dots,J$; and define $\bar{S}_{0,U}^2$ to be the counterpart of $S_{U}^2$ in the case of half-and-half sample splitting. Both $S_{U}^2$ and $\bar{S}_{0,U}^2$ can be calculated straightforwardly while implementing MCCV and its half-and-half sample splitting, similar to the calculation of $S_{R}^2$ and  $\bar{S}_{0}^2$. Based on the relationship between variance and sample variance as indicated by (\ref{samplevariance_rho}), we have 
\begin{equation*}
\frac{\mathbb{E}[\bar{S}_{0}^2]}{2\mathbb{E}[S_{R}^2]}\frac{\mathbb{E}[S_{U}^2]}{\mathbb{E}[\bar{S}_{0,U}^2]}=\frac{\text{Var}[\widehat{R}_{1}(D_1)]}{2\text{Var}[\widehat{R}_{1}]}\frac{1-\rho^{n/2}}{1-\rho}\frac{\text{Var}[U_i(D^{(1)}_{tr})]}{\text{Var}[U_i(D^{(1)}_{1,tr})]}\frac{1-\rho_2}{1-\rho_2^{n/2}},
\end{equation*}
which is slightly larger than the quantity $\text{Var}[\widehat{R}_{1}(D_1)]/(2\text{Var}[\widehat{R}_{1}])$ $\times\text{Var}[U_i(D^{(1)}_{tr})]/\text{Var}[U_i(D^{(1)}_{1,tr})]$ as $\rho^{n/2}<\rho$ and $\rho_2^{n/2}<\rho_2$, but should be quite close to $\text{Var}[\widehat{R}_{1}(D_1)]/(2\text{Var}[\widehat{R}_{1}])$ when all of the correlations $\rho^{n/2}$, $\rho$, $\rho_2^{n/2}$ and $\rho_2$ are close to zero. Approximately, we use $\max\{1, \bar{S}_{0}^2S_{U}^2/(2S_{R}^2 \bar{S}_{0,U}^2)\}$ as the inflation factor in practice and define the adjusted estimate of $\rho$ as follows:
\begin{equation}\label{rho_adj}
    \hat{\rho}^{\text{adj}}=\max\{1, \bar{S}_{0}^2S_{U}^2/(2S_{R}^2 \bar{S}_{0,U}^2)\}\hat{\rho}^{n/2}.
\end{equation}

\section{Model selection for dynamic treatment regimes}
In this section, we adapt the model selection method developed in Section \ref{sec:model_selection_single} to the context of  estimating optimal dynamic treatment regimes (DTRs), where a sequence of treatment decision rules tailored to the evolving health status is required to be determined. To this end, we firstly  generalize the notations  introduced in Section \ref{sec:model_selection_single} to the setting of multiple decision points and set up the framework of optimal DTRs.
\subsection{Notations and framework}
Suppose the follow-up records, for each individual $i$, are collected in time order as $(L_{1i}, A_{1i},\dots,L_{Ki}, A_{Ki},Y_i)$, where $A_{ki}$ is the observed treatment assignment, at the $k$th decision point, taking a value in $\{0,1\}$; $L_{ki}$ is the vector of covariates (also referred to as the intermediate outcome) collected after $A_{k-1,i}$ but prior to $A_{ki}$; and the definition of $Y_i$ is the same as before, namely, the outcome of interest observed at the end of the follow-up study, with larger values preferred. Moreover, we denote the covariate and treatment history available before the $k$th decision point by $H_k$, i.e., $H_k=(L_1,A_1,\dots,L_{k-1},A_{k-1},L_k)$. As for the counterfactual framework, we introduce $L^{\ast}_{k}(\bar{a}_{k-1})$ for $k=2,\dots,K$ and $Y^{\ast}(\bar{a})$ as the potential intermediate and final outcomes that would have been observed had the treatment history $\bar{a}_{k-1}=(a_1,\dots,a_{k-1})$ and $\bar{a}:=\bar{a}_K$ been followed. As there is no treatment assignment before $L_1$, $L_1^{\ast}$ is exactly the same as $L_1$. Let $\bar{\mathcal{A}}_{k}=\mathcal{A}_{1}\times \cdots \times \mathcal{A}_{k}$ be the set of feasible treatment histories up to and including the decision $k$ for $k=1,\dots,K$ with $\bar{\mathcal{A}}:=\bar{\mathcal{A}}_{K}$,  and define $O^{\ast}=\{L^{\ast}_1,L^{\ast}_2(a_1),\dots, L^*_{K}(\bar{a}_{K-1}),Y^{\ast}(\bar{a}):   \bar{a}\in \bar{\mathcal{A}}\}$ to be the collection of all potential outcomes. In general, for any given vector $v$ with $K$ elements, we use $\bar{v}_k$ to denote the elements up to and including the $k$th, i.e, $\bar{v}_k=(v_1,\dots,v_k)$, and use $\underline{v}_k$ to denote the future (including the current) elements, i.e., $\underline{v}_k=(v_k,\dots,v_K)$.

A DTR $\text{\slshape g}=(\text{\slshape g}_1,\dots,\text{\slshape g}_K)$ is a sequence of decision rules, one per decision point. The $k$th rule $\text{\slshape g}_k:=\text{\slshape g}_{k}(h_k)$ is a deterministic function, taking  the history $h_k$ prior to the decision point $k$ as input and outputs a treatment option $a_{k}\in \mathcal{A}_k$ for $k=1,\dots,K$. Let $\mathcal{G}$ denote the class of well-defined dynamic treatment regimes. The optimal DTR is defined as the one which maximizes the expected potential final outcome over $\mathcal{G}$. That is
\begin{equation}\label{optimal_DTR}
\begin{aligned}
\text{\slshape g}^{\text{opt}}&=\arg \max_{\text{\slshape g} \in \mathcal{G}} \mathbb{E} [Y^{\ast}(\text{\slshape g})]\\
&= \arg \max_{\text{\slshape g} \in \mathcal{G}} \mathbb{E}\left[Y^{\ast}\left(a_1=\text{\slshape g}_1(L^{\ast}_1),\dots,a_K=\text{\slshape g}_K\left(\bar{L}^{\ast}_{K}(\bar{a}_{K-1}),\bar{a}_{K-1}\right)\right)\right],
\end{aligned}
\end{equation}
where $Y^{\ast}(\text{\slshape g})$, defined by the second line of (\ref{optimal_DTR}), is the potential final outcome had the regime $\text{\slshape g}$ been followed, and the expectation is in fact taken with respect to the joint distribution of $O^{\ast}_{\text{\slshape g}}=(L^{\ast}_1,L^{\ast}_2(\text{\slshape g}_1),\dots, L^*_{K}(\bar{\text{\slshape g}}_{K-1}),Y^{\ast}(\text{\slshape g}))$. When $K=1$, we have $\text{\slshape g}^{\text{opt}}=\arg \max_{\text{\slshape g} \in \mathcal{G}} \mathbb{E} [Y^{\ast}(\text{\slshape g}(L))]=\arg \max_{\text{\slshape g} \in \mathcal{G}} \mathbb{E}\{\mathbb{E} [Y^{\ast}(\text{\slshape g}(L))|L]\}$, which under the assumptions given in Section \ref{sec:model_selection_single}, can be expressed in terms of the observed data $(L,A,Y)$, that is 
\begin{equation*}
\begin{aligned}
\mathbb{E}\{\mathbb{E} [Y^{\ast}(\text{\slshape g}(L))|L]\}=\mathbb{E}\{\mathbb{E} [Y|L,A=\text{\slshape g}(L)]\}.
\end{aligned}
\end{equation*}
If the treatment effect function $\tau$ is known,  the decision rule given by $\text{\slshape g}(L)=I\{\tau(L)>0\}$ is obviously the optimal one. When $K>1$, the identification of $\text{\slshape g}^{\text{opt}}$ requires stronger assumptions: (consistency) $L_k^{\ast}(\bar{a}_{k-1})=L_k$ if $\bar{A}_{k-1}=\bar{a}_{k=1}$, for any $\bar{a}_{k-1} \in \bar{\mathcal{A}}_{k-1}$, $k=2,\dots,K$ , and $Y^{\ast}(\bar{a})=Y$ if $\bar{A}=\bar{a}$;  (no unmeasured confounders) $A_k \ind O^{\ast} |H_k$ for $k=1,\dots K$; and the stable unit treatment value assumption remains the same. 
These assumptions establish the equivalence between the conditional distribution of the potential
outcomes and that of the observed outcomes, allowing the estimation of $\text{\slshape g}^{\text{opt}}$ from observed data by introducing statistical models. Generally speaking, $\text{\slshape g}^{\text{opt}}=(\text{\slshape g}^{\text{opt}}_1,\dots,\text{\slshape g}^{\text{opt}}_K)$ can either be estimated simultaneously in one step by optimizing the corresponding objective function derived from (\ref{optimal_DTR}), or be learned sequentially via backward induction, starting from the $K$th decision to obtain $\text{\slshape g}^{\text{opt}}_K$ and then moving backwards until $k=1$. The backward induction approach, compared with the simultaneous approach, is less computationally intensive as a single decision rule is to be determined in each stage, but more sensitive to model misspecification since the impact of a poor estimation of $\text{\slshape g}^{\text{opt}}_k$ in the $k$th decision point will propagate over later considered stages (the $k-1,\dots,1$ earlier decision points). Although A-learning mitigates model misspecification by specifying the treatment-related term only and leaving the treatment-free term unspecified,   model selection on the treatment-related term is still helpful in improving the final performance of the estimated optimal DTR, especially when the candidate models are restricted to be clinically interpretable. 

\subsection{A-learning with model selection}
Model selection for CATE developed in Section \ref{sec:model_selection_single} can be naturally generalized to A-learning by considering the following contrast functions
\begin{equation}\label{contrast}
    C_{k}(h_k)=\mathbb{E}[Y^{\ast}(\bar{a}_{k-1},1,\underline{\text{\slshape g}}_{k+1}^{\text{opt}})-Y^{\ast}(\bar{a}_{k-1},0,\underline{\text{\slshape g}}_{k+1}^{\text{opt}})|H_{k}=h_k],\quad  k=1,\dots,K,
\end{equation}
which characterize the effect of being treated at the $k$th point on the final outcome when the optimal treatment rules are followed from $k+1$ onwards. Obviously, when $K=1$, the contrast function  $C(h)$ is exactly $\tau(l)$. Similar to the case of CATE where a surrogate for $\tau$ is required, we now need to construct a surrogate for the contrast function $C_{k}(h_k)$ based on $\{(H_{k,i},A_{k,i},\widehat{V}_{k+1,i})\}_{i=1}^n$, where $\widehat{V}_{k+1}$ is an estimate of the value function $V_{k+1}:=V_{k+1}(\bar{l}_{k+1},\bar{a}_{k})$ and
$$V_{k+1}(\bar{l}_{k+1},\bar{a}_{k})=\mathbb{E}[Y^{\ast}(\bar{a}_{k},\underline{\text{\slshape g}}^{\text{opt}}_{k+1})|\bar{L}^{\ast}_{k+1}(\bar{a}_{k})=\bar{l}_{k+1}]=\mathbb{E}[Y^{\ast}(\bar{a}_{k},\underline{\text{\slshape g}}^{\text{opt}}_{k+1})|\bar{L}_{k+1}=\bar{l}_{k+1},\bar{A}_{k}=\bar{a}_{k}],$$
for $k=1,\dots, K-1$, with $V_{K+1}=Y$. 

Given the estimated contrast functions from $k+1$ to $K$, $\widehat{V}_{k+1}$ in A-learning is obtained as follows:
\begin{equation}\label{response_A}
\begin{aligned}
\widehat{V}_{k+1}=Y+\sum_{m=k+1}^{K}\left(I\left\{\widehat{C}_m(H_m)>0\right\}-A_m\right)\widehat{C}_m(H_m),
\end{aligned}
\end{equation}
which then serves as the response variable guiding the estimation and validation of $C_k$. Specifically, $\widetilde{V}_{k+1,i}$, the {\emph{surrogate}} for  $C_k(H_{k,i})$, is constructed by
\begin{equation}\label{surrogate_A}
\begin{aligned}
\widetilde{V}_{k+1,i}=(2A_{k,i}-1)(\widehat{V}_{k+1,i}-\widehat{V}_{k+1,i^{\prime}}),
\end{aligned}
\end{equation}
where $i^{\prime}$ is the most similar individual (with respect to $H_k$) in the opposite treatment group to individual $i$ for $i,i^{\prime}\in I_{val}$. In this way, the calculation of MCCV and its variance developed earlier for CATE can be directly applied to perform model selection for $C_k$ supplemented with the estimated variance to indicate the confidence in the selection.

In summary,  A-learning with model selection on the contrast functions proceeds as follows: starting from the $K$th decision point, the MCCV procedure is applied to $D=\{(H_{K,i},A_{K,i},Y_{i})\}_{i=1}^n$ to obtain $\widehat{R}_{cv,K}$ for each candidate model separately or for the difference in the generalization error between two models of interest, depending on the definition of $U_i(D_{tr})$. A selection can be made solely on the basis of $\widehat{R}_{cv,K}$'s or by combining $\widehat{R}_{cv,K}$ with its variance. The latter can be carried out using hypothesis testing especially when there is a particular model that is to be {\emph{protected}} due to, for example, its better clinical interpretability. Once a model for $C_{K}$ has been selected, we fit the selected model to the whole data set to get the estimated contrast function $\widehat{C}_{K}$ and derive the optimal treatment rule  as $\hat{\text{\slshape g}}_{K}^{\text{opt}}(H_K)=I(\widehat{C}_K(H_K)>0)$. The value function $\widehat{V}_{K}$, calculated from (\ref{response_A}), then works as the updated response variable in stage $K-1$. For $k=K-1,\dots,1$, model selection for $C_{k}$ is performed on $D=\{(H_{k,i},A_{k,i},\widehat{V}_{k+1,i})\}_{i=1}^n$ with the surrogate $\widetilde{V}_{k+1,i}$ constructed from (\ref{surrogate_A}) for $i \in I_{val}$; and $\hat{\text{\slshape g}}_{k}^{\text{opt}}$ is derived from the selected and then estimated $\widehat{C}_{k}$. The backward induction procedure, as outlined in Algorithm \ref{algorithm}, finally outputs the estimated optimal dynamic treatment regime $\hat{\text{\slshape g}}^{\text{opt}}=(\hat{\text{\slshape g}}^{\text{opt}}_1,\dots,\hat{\text{\slshape g}}^{\text{opt}}_K)$.

\begin{algorithm}[H]
	\caption{\label{algorithm} A-learning with MCCV} 
	\begin{algorithmic}
            \State Initialize $\widehat{V}_{K+1}:=Y$ and the threshold of $p$-value, $p_0<0.5$, if there is preference among candidate models.
		\For {$k=K,\ldots,1$}
                \begin{itemize}
                     
                    \item[1] Apply the MCCV procedure shown in Figure \ref{fig:mccv} to $D=\{(H_{k,i},A_{k,i},\widehat{V}_{k+1,i})\}_{i=1}^n$ and calculate $\widehat{R}_{cv,k}$ with $U_i(D_{tr})=[\widetilde{V}_{k+1,i}-\widehat{C}_{lin,k}(H_{k,i};D_{tr})]^2-[\widetilde{V}_{k+1,i}-\widehat{C}_{tre,k}(H_{k,i};D_{tr})]^2$
                    \item[2] Select the model/estimator 
                    \begin{itemize}
                        \item if no preference, choose the linear model if $\widehat{R}_{cv,k}\leq 0$ and the tree-based model otherwise;
                        \item if, for example, the linear model is preferred, 
                        \begin{itemize}
                            \item[2.1] calculate the variance of $\widehat{R}_{cv,k}$, via data-splitting;
                            \item[2.2] calculate the $p$-value $\hat{p}=1-\Phi\left(\widehat{R}_{cv,k}/\sqrt{\widehat{\text{Var}}(\widehat{R}_{cv,k})}\right)$, and choose the tree-based model if $\hat{p}<p_0$ 
                        \end{itemize}
                    \end{itemize}
                    
                    \item [3] Fit the selected model on the original data to get the estimated contrast function $\widehat{C}_k$ and derive the optimal treatment rule at decision $k$: $\hat{\text{\slshape g}}_{k}^{\text{opt}}(H_k)=I(\widehat{C}_k(H_k)>0)$. 
                    \item [4] $\widehat{V}_{k}=\widehat{V}_{k+1}+\left(\hat{\text{\slshape g}}_{k}^{\text{opt}}(H_k)-A_k\right)\widehat{C}_k(H_k)$
           
           \end{itemize}

		\EndFor
            
	\end{algorithmic} 
        \Return $\hat{\text{\slshape g}}^{\text{opt}}=(\hat{\text{\slshape g}}^{\text{opt}}_1,\dots,\hat{\text{\slshape g}}^{\text{opt}}_K)$
\end{algorithm}

\section{Simulation}\label{sec:simulation}
We conduct simulation studies to demonstrate the performance of our proposed variance estimator of $\widehat{R}_{cv}$ in Section \ref{sec:sim_var} and the improved performance of $\hat{\text{\slshape g}}^{\text{opt}}$ in Section \ref{sec:sim_dtr},  in the sense of the value function $\mathbb{E}[Y^*(\hat{\text{\slshape g}}^{\text{opt}})]$ and the decision accuracy, when model selection is performed while determining the treatment decision rules. For ease of illustration, the candidate models/estimators are set as the linear and tree-based models shown in Figure \ref{fig:mse_samplesize}. Specifically, the linear model for $C_{k}$ is specified as 
\begin{equation}\label{linear_mod}
C_{lin,k}(H_k):=C_k(H_k;\alpha_k,\beta_k)=\alpha_{k}+\beta_{k}^{\top}L_k,
\end{equation}
where only the current biomarker status $L_k$ is included in the linear form for simplicity and the derived decision boundary is a hyperplane defined by $\alpha_{k}+\beta_{k}^{\top}L_k=0$, which reduces to a line  when $L_k$ is two-dimensional, as in Figure \ref{fig:mse_samplesize}. The tree-based model for $C_k$, specified by the tree structure $\Pi_k=\{\ell_{k,1},\dots, \ell_{k,\#\Pi_k}\}$ and the leaf-wise treatment effects $\gamma_k$,  is of the following form:
\begin{equation}\label{tree_mod}
C_{tree,k}(H_k):=C_k(H_k;\gamma_k,\Pi_k)=\sum_{m=1}^{\#\Pi_k}\gamma_{k,m}I\{L_k\in \ell_{k,m}\},
\end{equation}
where each leaf $\ell_{k,m}$ represents a rectangular region in the space of $L_k$ and the tree structure $\Pi_k$ indicates a partition of the space, leading to the boundaries which are parallel to the coordinate axes. In contrast to $C_{lin,k}$ where a linear structure is pre-specified,  $C_{tree,k}$ allows the model structure, i.e., the tree structure, to be learned from the data. Therefore, the tree-based model is data-adaptive but also more sensitive to the sample size and  data perturbation, as indicated by the subfigures (a)-(c) in Figure \ref{fig:mse_samplesize}. Due to such a difference in statistical properties between the linear and tree-based models, it is expected that their relative performance, in terms of the generalization error, in approximating the contrast function (the treatment effect function when $K=1$), depends on the available sample size and the profile of the true contrast function. 

Subfigures (d)-(f) in Figure \ref{fig:mse_samplesize} depict three scenarios for the true treatment effect function $\tau$, which are designed to favor one model over the other by controlling the steepness parameter in the logistic function of each dimension of $L$. Specifically, the true CATE $\tau$ is of the form 
\begin{equation}\label{true_tau}
  \tau(l):=\tau(l;c,s,z)=c(1-\zeta_1(l_1;s)\zeta_2(l_2;s)-z),
\end{equation}
where $\zeta_1(l_1;s)=1/[1+\exp\{s(l_1-20)\}]$ and $\zeta_2(l_2;s)=1/[1+\exp\{s(l_2-12)\}]$ are  logistic functions with a common steepness parameter $s$, $c$ is the scale parameter with $c=10$ in subfigures (e) and (f) and $c=30$ in subfigure (d), and $z$ controls the location of the boundary with $z=0.5$ in subfigures (e) and (f) and $z=0.75$ in subfigure (d). By setting $s=0.1$,  $\tau$ in subfigure (d) manifests a gradual change with $(l_1,l_2)$ and favors the linear model  whereas $\tau$ in subfigure (f) favors the tree-based model and exhibits a sharp boundary by increasing $s$ to 1. $\tau$ in subfigure (e) with $s=0.5$ lies in between; and model selection between the linear model and the tree-based model in this scenario should involve large uncertainty. Moreover, Figure \ref{fig:mse_samplesize} will serve as a reference throughout the simulation section since the $\tau$-risk for both models under each given sample size can be read from it.
For computation, we estimate $(\alpha_k,\beta_k)$ in (\ref{linear_mod}) via g-estimation with the R package \textbf{DTRreg} \citep{DTRreg} while  $(\gamma_k,\Pi_k)$ in (\ref{tree_mod}) are estimated with the causal tree method developed in the R package \textbf{htetree} \citep{htetree}.

\subsection{Estimation of $\text{Var}[\widehat{R}_{cv}]$} \label{sec:sim_var}
In this section, we compare the performance of our proposed estimator of $\text{Var}[\widehat{R}_{cv}]$ in Section~\ref{sec:var_mccv} with other existing methods. Data are generated as follows: $W \sim N_{(10,\infty)}(45,10^2)$, $L_{1}\sim N_{(0,\infty)}(20,5^2)$, $L_{2}\sim N_{(0,\infty)}(10,3^2)$ with $L=(L_1,L_2)$,
$A|W,L\sim \text{Bin}(1,p)$ with $p=\text{expit}(-2+0.05W)$, and 
$$Y|W,L \sim  N(100,2^2)-(I\{\tau(L)>0\}-A)\tau(L),$$
where $\tau$ takes the form of that in (\ref{true_tau}) with parameters set to three different configurations to give rise to the corresponding profiles shown in (d)-(f) in Figure \ref{fig:mse_samplesize}.

Table \ref{tab:var} summarizes the results obtained from different estimation methods under different sample sizes and  profiles of $\tau$. In addition to the proposed estimators $\text{var}_{\hat{\rho}^{n/2}}$ and  $\text{var}_{\hat{\rho}^{\text{adj}}}$ derived from $\hat{\rho}^{n/2}$ in (\ref{rho_n/2}) and $\hat{\rho}^{\text{adj}}$ in (\ref{rho_adj}) respectively, we also present the performance of $\text{var}_{\rho=0}$ and $\text{var}_{\rho=q}$, the variance estimators obtained by setting $\rho$ in (\ref{variance_rho}) to be $0$ and the validation proportion $q = 0.2$, respectively. Moreover, $\text{var}_{1/2}$ and $\text{var}_{(n_1^{\prime},n_2)}$ in Table \ref{tab:var} correspond to the estimators based on $\bar{S}_{cv}^2$, the average of the sample variances of $\{\widehat{R}_{cv}(D_1^{(b)}),\widehat{R}_{cv}(D_2^{(b)})\}$ for $b=1,\dots, B$ from the half-and-half sample splitting, where $\widehat{R}_{cv}(D_1^{(b)})$'s in $\text{var}_{1/2}$ are constructed with the original proportion $q$, and $\widehat{R}_{cv}(D_1^{(b)})$'s in $\text{var}_{(n_1^{\prime},n_2)}$ are constructed with the proportion $2q$ so that the validation size $n_2$ aligns with that in $\widehat{R}_{cv}(D)$. The results in Table \ref{tab:var} are based on 1000 Monte Carlo repetitions and the Monte Carlo variance of $\widehat{R}_{cv}(D)$, denoted by $\text{var}^{\ast}$ in Table \ref{tab:var}, serves as the reference when assessing the performance of the variance estimators. 

As expected, $\text{var}_{\rho=0}$ significantly underestimates  $\text{Var}[\widehat{R}_{cv}]$ in all cases in Table \ref{tab:var} by ignoring the correlation among $\widehat{R}_1,\dots, \widehat{R}_J$ with $J=100$ in the simulation studies. In contrast,  $\text{var}_{\rho=q}$ overestimates $\text{Var}[\widehat{R}_{cv}]$ since $\rho$ in CATE under our specified setting is much smaller than $q$, as indicated by $\hat{\rho}^{n/2}$ and $\hat{\rho}^{\text{adj}}$ in Table \ref{tab:var}. Both $\text{var}_{1/2}$ and $\text{var}_{(n_1^{\prime},n_2)}$ overestimate $\text{Var}[\widehat{R}_{cv}]$ due to the decreased training size $n_1^{\prime}$, the increased $\rho$ in $\text{var}_{(n_1^{\prime},n_2)}$ and the decreased validation size (also the decreased training size) in $\text{var}_{1/2}$. Our proposed estimators $\text{var}_{\hat{\rho}^{n/2}}$ and  $\text{var}_{\hat{\rho}^{\text{adj}}}$ generally provide reasonable approximations to $\text{Var}[\widehat{R}_{cv}]$. The results on $\hat{\rho}^{n/2}$ in Table \ref{tab:var} show that the correlation among $\widehat{R}_1,\dots, \widehat{R}_{J}$ tends to increase with the sample size especially under settings (e) and (f) where the performance of the data-adaptive model, here the tree-based model, considerably gets better and stabilizes as the sample size doubles. In this case, $\text{var}_{\hat{\rho}^{n/2}}$ derived from $\hat{\rho}^{n/2}$ tends to underestimate $\text{Var}[\widehat{R}_{cv}]$. By applying the inflation factor to $\hat{\rho}^{n/2}$, $\text{var}_{\hat{\rho}^{\text{adj}}}$ based on $\hat{\rho}^{\text{adj}}$  provides a better approximation to $\text{Var}[\widehat{R}_{cv}]$ under settings (e) and (f) but slightly overestimates  $\text{Var}[\widehat{R}_{cv}]$ under the setting (d).

$\widehat{R}_{cv}$ itself provides a point estimate of $\mathbb{E}[R(\hat{\tau}_{lin})-R(\hat{\tau}_{tre})]$, the difference between the linear model and the tree-based model in the mean of $\tau$-risk under a given sample size $n(1-q)$. This estimate is a bit different from the value obtained under size $n$, indicated by the solid points in (a)-(c) in Figure \ref{fig:mse_samplesize}.  The error bars in subfigures (a)-(c) reflect the uncertainty of $R(\hat{\tau})$ ($\hat{\tau}$ can be $\hat{\tau}_{lin}$ or $\hat{\tau}_{tre}$), and is related to but different from the uncertainty of $\widehat{R}_{cv}$, since the latter involves the variation in assessing $\hat{\tau}$ (equivalently, the variation in validation sets) additional to the variation in estimating $\hat{\tau}$ (equivalently, the variation in training sets). In other words, the variance of $R(\hat{\tau})$ is determined by the nature of $\hat{\tau}$ while the variance of $\widehat{R}_{cv}$ also depends on the cross-validation method employed to estimate the risk. The average of $\widehat{R}_{cv}$ over 1000 Monte Carlo repetitions is reported in Table \ref{tab:var}, denoted by $\text{av}(\widehat{R}_{cv})$. As can be seen from  $\text{av}(\widehat{R}_{cv})$, the linear model outperforms the tree-based model across different sample sizes under the setting (d) while the opposite conclusion is obtained under the setting (f); consistent with the results shown in (a) and (c) in Figure \ref{fig:mse_samplesize}. Despite $\widehat{R}_{cv}<0$ when $n=200$ under the setting (f), the variance of $\widehat{R}_{cv}$ implies that there is no significant difference between the two models in this case. The conclusion on model selection under the setting (e) varies with the sample size and the crossing happens around $n=1000$. Model selection in practice can be performed solely based on $\widehat{R}_{cv}$ or by combining $\widehat{R}_{cv}$ with its estimated variance if there is a preference among candidate models. 

\begin{table}[H]
{\small
\centering
\caption{\label{tab:var} The performance of different methods in estimating $\text{Var}[\widehat{R}_{cv}]$ under three settings with the Monte Carlo variance `$\text{var}^{*}$' serving as the reference.} \vspace{1mm}
\begin{tabular}{crrrrrrrrr@{\hskip 0.005in}r}
  \hline
 $n$& \multicolumn{1}{c}{$\text{av}(\widehat{R}_{cv})$}& \multicolumn{1}{c}{$\text{var}^{*}$} & \multicolumn{1}{c}{$\hat{\rho}^{n/2}$}
   &\multicolumn{1}{c}{$\text{var}_{\hat{\rho}^{n/2}}$}&  \multicolumn{1}{c}{$\hat{\rho}^{\text{adj}}$}& \multicolumn{1}{c}{ $\text{var}_{\hat{\rho}^{\text{adj}}}$} &\multicolumn{1}{c}{$\text{var}_{\rho=0}$}&\multicolumn{1}{c}{$\text{var}_{\rho=q}$} &\multicolumn{1}{c}{$\text{var}_{1/2}$}&\multicolumn{1}{c}{$\text{var}_{(n_1^{\prime},n_2)}$}\\ 
  \hline \\
  & \multicolumn{9}{c}{(a) and (d): $s=0.1$} \vspace{1.5mm}\\
  $200$  & $-2.2293$   &0.4224  & 0.05457 &0.4258 &  0.0585 & 0.4486 & 0.0620 &1.6131& 1.5247 & 1.9348\\ 
  $500$  & $-1.6878$ &  0.1131  & 0.0557 & 0.1094 &0.0599 &0.1155& 0.0159& 0.4131 &0.3371 &0.4060\\ 
  $1000$ & $-1.3198$ & 0.0387  & 0.0548& 0.0402 &0.0584 &0.0422&0.0059 & 0.1536 &0.1082 & 0.1420\\ \vspace{1.5mm}
  $2000$ & $-0.9776$ &0.0145 &0.0537 & 0.0148 & 0.0571& 0.0154 &0.0022 & 0.0565 &0.0393 &  0.0486\\ 
  & \multicolumn{9}{c}{(b) and (e): $s=0.5$} \vspace{1.5mm}\\
  $200$  & $-1.8631$  &0.5066 & 0.0469 & 0.4544 & 0.0553 &0.4910 & 0.0757& 1.9685& 1.3884  &  2.2091\\ 
  $500$  & $-0.8014$ &0.2047 & 0.0606 & 0.1591 & 0.0755 &0.1889 &0.0212 & 0.5522 & 0.4016 &0.5248 \\ 
  $1000$  &$-0.2002$ &0.1025 & 0.0809 & 0.0918 & 0.0819
  &0.0997 &0.0093 & 0.2427 & 0.2128 &0.2602\\ \vspace{1.5mm}
  $2000$  & 0.2517 &0.0529 & 0.0959 & 0.0486  & 0.1073 &0.0517 & 0.0041 & 0.1062 & 0.1065 & 0.1269 \\ 
  & \multicolumn{9}{c}{(c) and (f): $s=1$} \vspace{1.5mm}\\
  $200$  & $-1.0409$ &0.9542 & 0.0496 &0.8037 & 0.0657& 0.9237 &0.1283 &3.3155 & 2.3106 & 3.7248\\ 
  $500$  & $0.9742$ &0.4536 & 0.0715 & 0.3584 & 0.0807& 0.4114 & 
  0.0404 & 1.0505 & 0.8174& 0.9989\\ 
  $1000$  & 1.9214 &0.2584 & 0.0952 & 0.2266 & 0.0991& 0.2438 & 0.0196 &0.5104 & 0.4827 & 0.5855\\ 
  $2000$  & 2.5743 &0.1373 & 0.1118 &0.1275 & 0.1216 & 0.1370 & 0.0094 & 0.2437 & 0.2649 & 0.3012\\ 

\hline
\end{tabular}
\label{tab:sim1_res}}
\end{table}

\subsection{Model selection in DTR}\label{sec:sim_dtr}
We consider DTR with two decision points ($K=2$) and generate data as follows:
$W \sim N_{(10,80)}(45,10^2)$, $L_{11}\sim N_{(0,40)}(20,5^2)$, $L_{12}\sim N_{(0,30)}(10,3^2)$ with $L_1=(L_{11},L_{12})$,
$A_1|H_1\sim \text{Bin}(1,p_1)$ with $p_1=\text{expit}(-2+0.05W)$; $L_{21}|H_1,A_1\sim N(L_{11},3^2)$, $L_{22}|H_1,A_1\sim N(L_{12},2^2)$ with $L_2=(L_{21},L_{22})$, 
$A_2|\bar{L}_2,A_1\sim \text{Bin}(1,p_2)~\text{with}~p_2=\text{expit}(-1+0.04(L_{21}+L_{22}));$ and 
$$Y|\bar{L}_2,\bar{A}_2 \sim N(100,2^2)-(I\{C_1(L_1)>0\}-A_1)C_1(L_1)-(I\{C_2(L_2)>0\}-A_2)C_2(L_2),$$
where the true contrast functions $C_k(L_k):=C_k(L_k;c_k,s_k,z_k)$ for $k=1,2$ are of the form given in (\ref{true_tau}) and the true optimal DTR is $(\text{\slshape g}^{\text{opt}}_1, \text{\slshape g}^{\text{opt}}_2)=(I\{C_1(L_1)>0\}, I\{C_2(L_2)>0\})$. Table \ref{tab:sim_dtr_setting} summaries the simulation settings we consider in this section, where `true model' in the header refers to the models which should be selected in stage 1 and stage 2. For example, the linear model should be selected in stage 1 and the tree-based model should be selected in stage 2 in case iii, since the steepness parameters are set to $(s_1,s_2)=(0.1,1)$. Additionally, the models underlined in case i and ii in the table correspond to the situation where the variation of the $\tau$-risk is large enough that there is no significantly better model although the underlined models are better in terms of the mean of $\tau$-risk. The sample size is fixed to be 1000 across all cases.

\begin{table}[H]
\centering
\caption{\label{tab:sim_dtr_setting}Simulation settings in Section \ref{sec:sim_dtr} with $n=1000$.} \vspace{1mm}
\begin{tabular}{c c r r r r r c }
\hline 
\multicolumn{1}{c}{case}&\multicolumn{1}{c}{$c_1$}  & \multicolumn{1}{c}{$s_1$} & \multicolumn{1}{c}{$z_1$}  & \multicolumn{1}{c}{$c_2$} & \multicolumn{1}{c}{$s_2$} & \multicolumn{1}{c}{$z_2$} &\multicolumn{1}{c}{true model} \\ \hline \vspace{-2mm} \\ 
i & 30 & 0.1 & 0.75 & 10 & 0.5 & 0.5 & (linear, \underline{linear}) \\
ii & 10 & 0.5 & 0.5 & 30 & 0.1 & 0.75 &(\underline{linear}, linear) \\
iii & 30 & 0.1 & 0.75 & 30 & 1.0 & 0.5 & (linear, tree)\\
iv & 30 & 1.0 & 0.5 & 30 & 0.1 & 0.75 & (tree, linear)\\
\hline
\end{tabular}
\end{table}

\begin{figure}[H]
    \centering
       \subfigure[]{\includegraphics[width=0.45\textwidth, height=5.5cm]{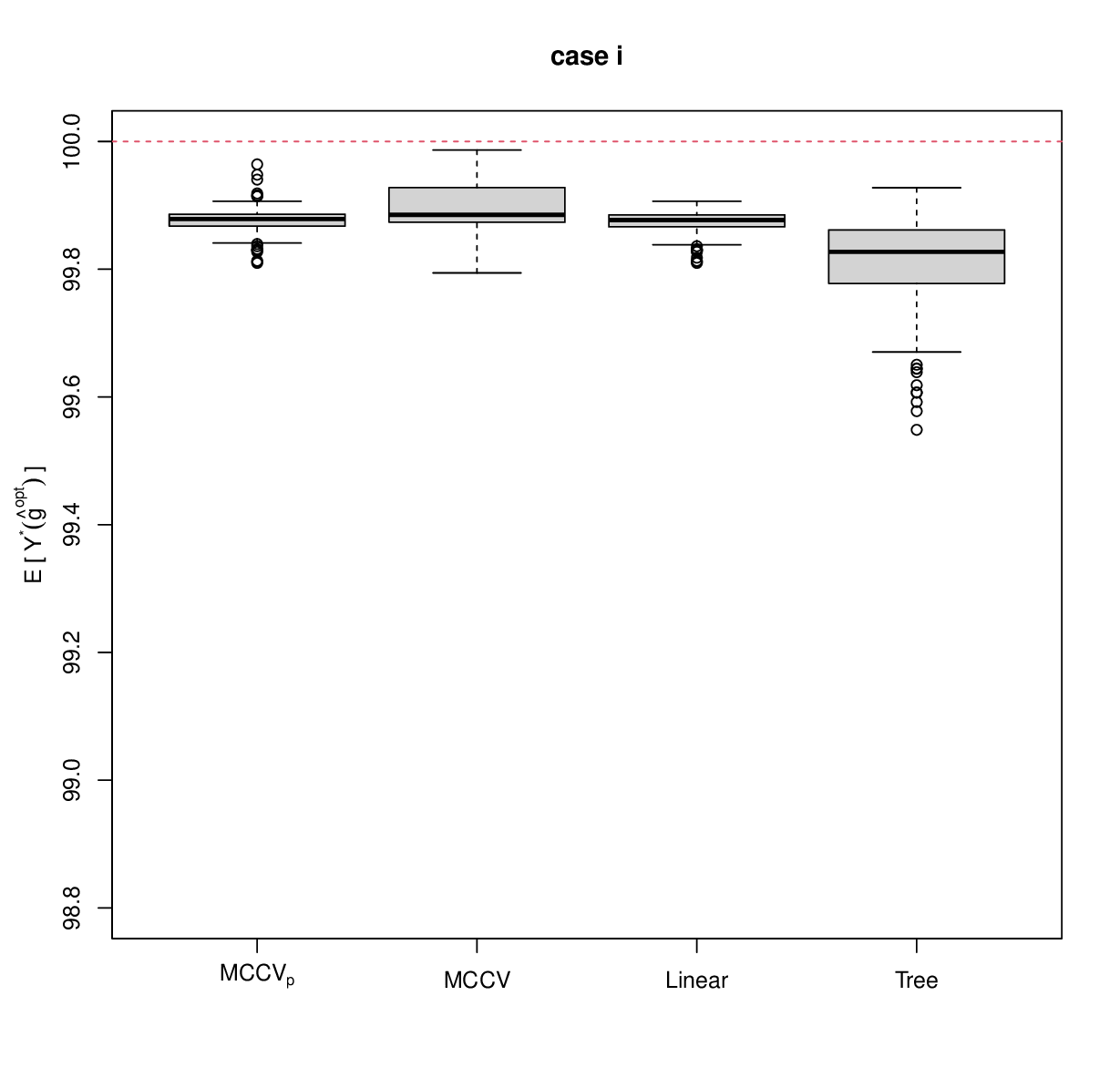}} 
    \subfigure[]{\includegraphics[width=0.45\textwidth, height=5.5cm]{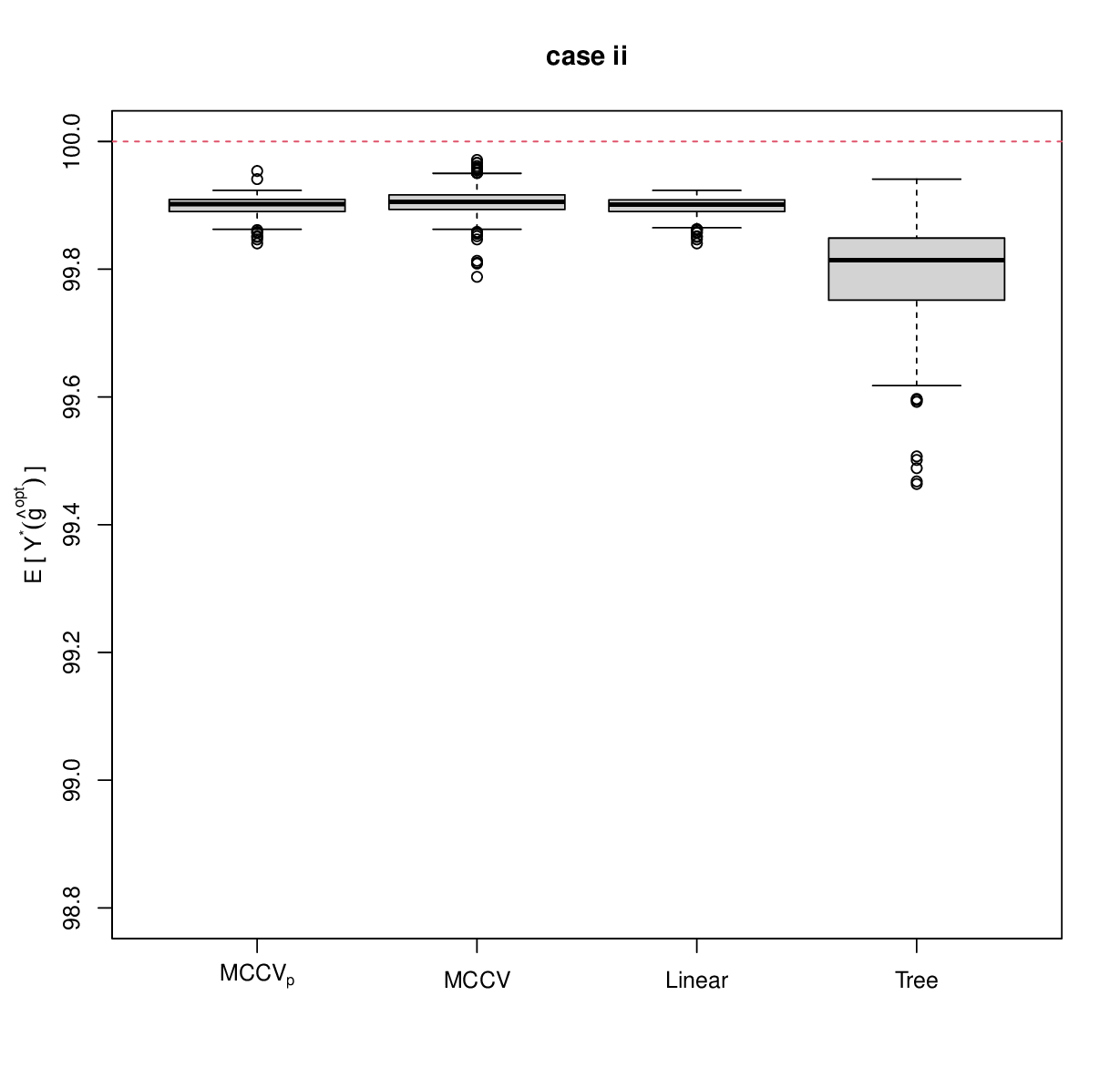}} \\
    \subfigure[]{\includegraphics[width=0.45\textwidth,height=5.5cm]{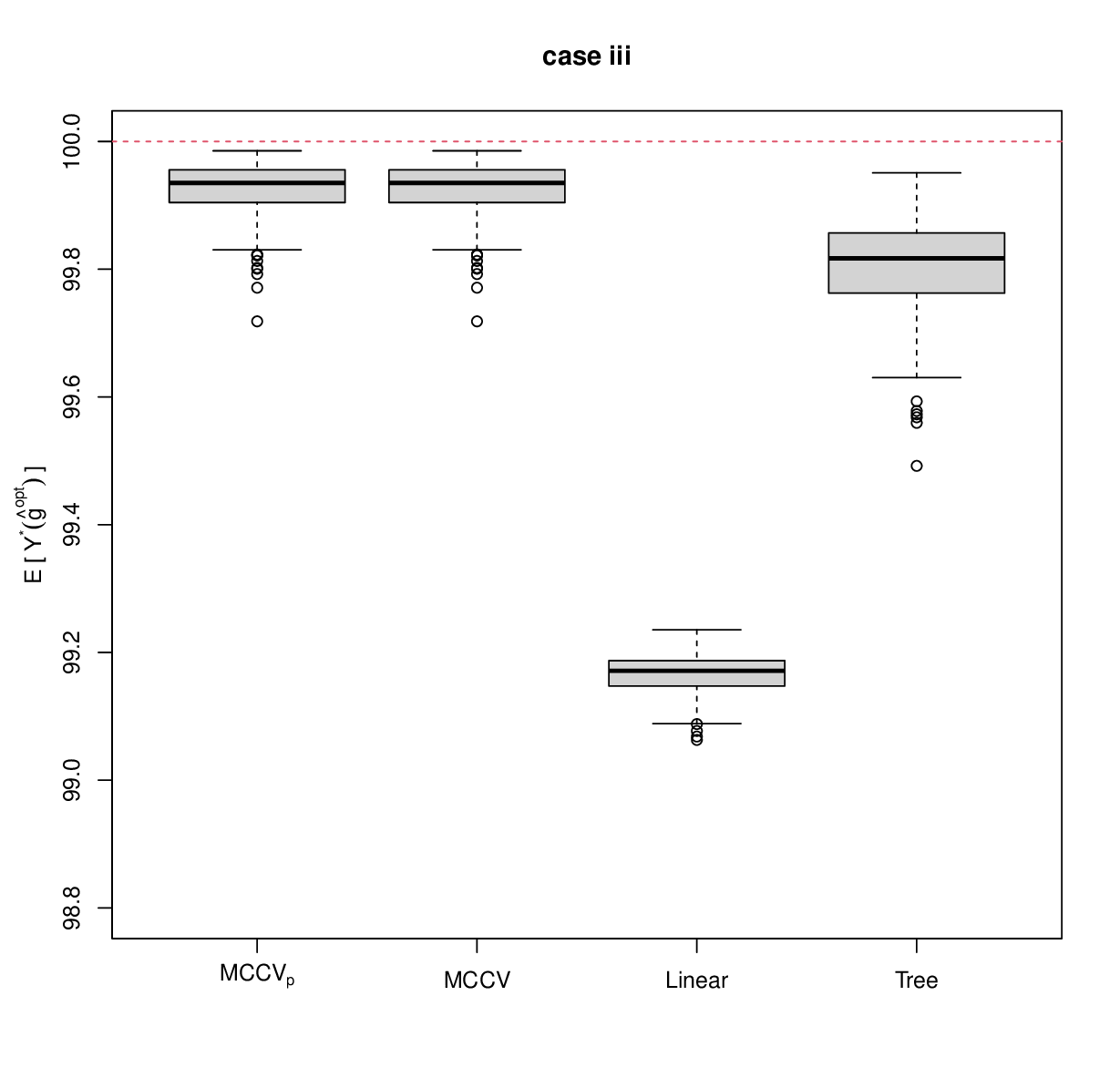}} 
     \subfigure[]{\includegraphics[width=0.45\textwidth,height=5.5cm]{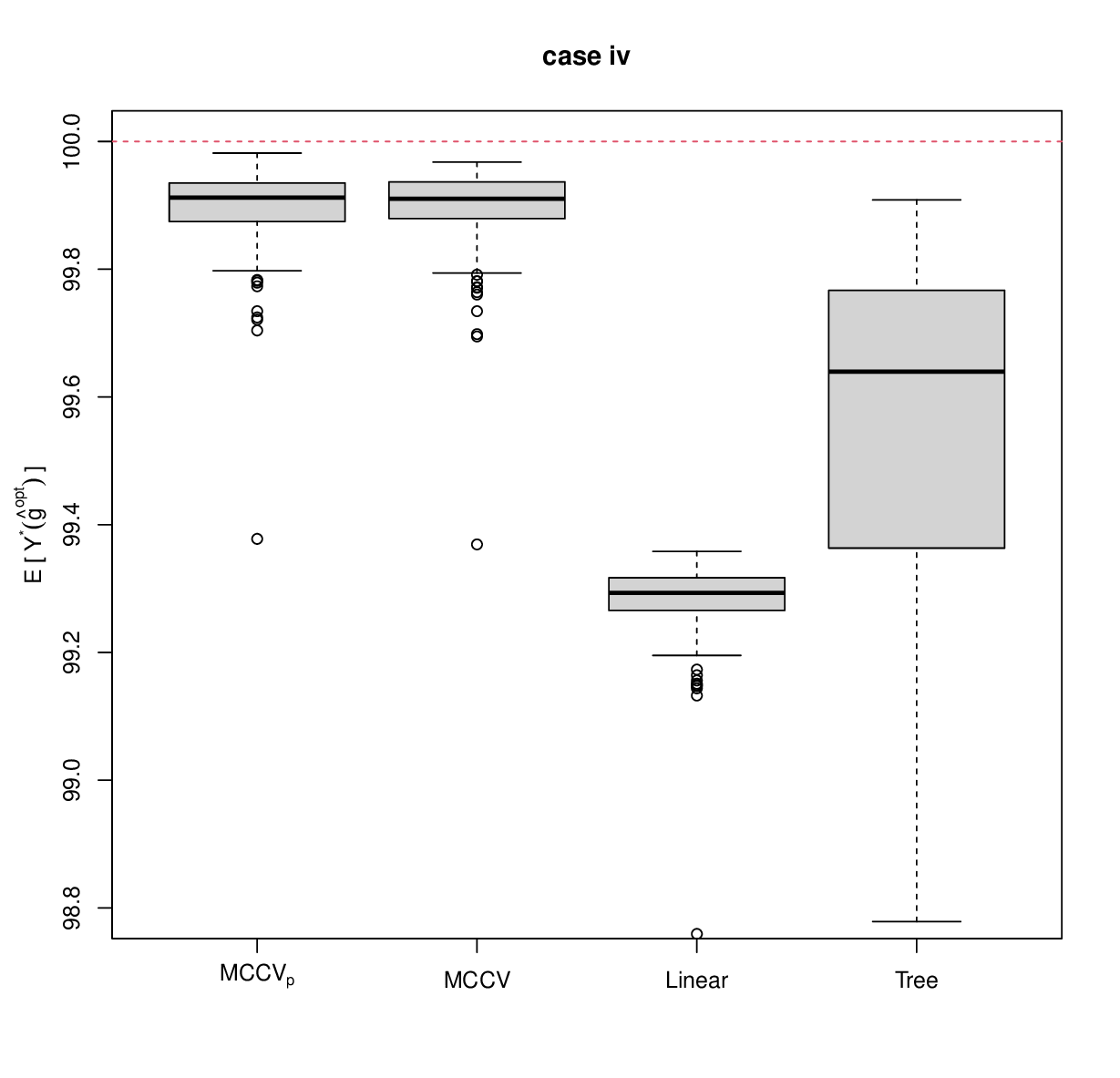}} 
    \caption{\label{fig:sim_dtr} Evaluation of $\mathbb{E} [Y^{\ast}(\hat{\text{\slshape g}}^{\text{opt}})]$ via Monte Carlo sampling with $\hat{\text{\slshape g}}^{\text{opt}}$ obtained from different approaches.}
    \end{figure}

We compare the performance of the following four approaches in estimating the optimal DTR: `Linear' (`Tree') refers to the approach that no model selection is performed and the linear (tree-based) model $C_{lin,k}$ ($C_{tree,k}$) for $k=1,2$ is automatically employed in both stages;  `MCCV' performs model selection solely based on $\widehat{R}_{cv}$ in each stage; and `$\text{MCCV}_p$' combines $\widehat{R}_{cv}$ with its estimated variance (here we use $\text{var}_{\hat{\rho}^{\text{adj}}}$) to base model selection on the conclusion obtained from hypothesis testing with a prespecified $p$-value threshold $p_0$, as outlined in Algorithm \ref{algorithm}. We set $p_0=0.05$ in $\text{MCCV}_p$ and the null hypothesis in each stage is specified to protect the linear model. The performance of each method is evaluated in terms of the decision accuracy (Table \ref{tab:DTR_accuracy}) and the value of the estimated optimal DTR $\hat{\text{\slshape g}}^{\text{opt}}$, $\mathbb{E}[Y^{\ast}(\hat{\text{\slshape g}}^{\text{opt}})]$ (Figure \ref{fig:sim_dtr}). Specifically, for a given $\hat{\text{\slshape g}}^{\text{opt}}=(\hat{\text{\slshape g}}_1^{\text{opt}},\hat{\text{\slshape g}}_2^{\text{opt}})$, $\mathbb{E}[Y^{\ast}(\hat{\text{\slshape g}}^{\text{opt}})]$ is calculated as the sample mean of  $10^5$ iid samples from the distribution $(W,L_1,L^{\ast}_2(\hat{\text{\slshape g}}_1^{\text{opt}}),Y^{\ast}(\hat{\text{\slshape g}}^{\text{opt}}))$, and the decision accuracy for stage $k$
is defined as $\sum_{i=1}^{10^5}I\{\hat{\text{\slshape g}}_k^{\text{opt}}(H_{k,i})=\text{\slshape g}_k^{\text{opt}}(H_{k,i})\}/10^5$ for $k=1,2$ and $\sum_{i=1}^{10^5}I\{\hat{\text{\slshape g}}_1^{\text{opt}}(H_{1,i})=\text{\slshape g}_1^{\text{opt}}(H_{1,i}),\hat{\text{\slshape g}}_2^{\text{opt}}(H_{2,i})=\text{\slshape g}_2^{\text{opt}}(H_{2,i})\}/10^5$ for the accuracy over both stages.

As can be seen from Table \ref{tab:DTR_accuracy}, incorporating model selection in estimating optimal DTRs improves the decision accuracies, especially in cases iii and iv where both methods, Linear and Tree, suffer from model misspecification in one of two stages. Moreover, Tree performs much worse in case iv than that in case iii since the incorrect model specification in the second stage will then affect the estimation in stage 1. This also applies to $\mathbb{E}[Y^{\ast}(\hat{\text{\slshape g}}^{\text{opt}})]$ and leads to 
increased variation in $\mathbb{E}[Y^{\ast}(\hat{\text{\slshape g}}^{\text{opt}})]$, as shown in subfigures (c) and (d) in Figure \ref{fig:sim_dtr}. Since  the steepness parameters $(s_1,s_2)$ identify the better model (i.e., `true model' in Table \ref{tab:sim_dtr_setting}) with 100\% certainty in cases iii and iv, both $\text{MCCV}_p$ and MCCV select the correct model over all simulations, leading to exactly the same performance between the two methods. In other words, when one model is considerably better than another, the prior preference expressed through the null hypothesis in $\text{MCCV}_p$ will not change the final decision on selection.  In the cases of i and ii where the `true model' leads to substantial uncertainty, $\text{MCCV}_p$ ends up selecting the linear model more frequently than MCCV, as indicated by the percentage of times that the tree-based model is selected among 200 Monte Carlo replications in Table \ref{tab:DTR_accuracy}. This protection/preference towards the linear model in $\text{MCCV}_p$, as shown in Table \ref{tab:DTR_accuracy} and Figure \ref{fig:sim_dtr}, results in negligible loss in the mean values of decision accuracies and $\mathbb{E}[Y^{\ast}(\hat{\text{\slshape g}}^{\text{opt}})]$ but increased stability in these quantities, compared with MCCV. Moreover, Linear in fact serves as the reference in cases i and ii, and $\text{MCCV}_p$ achieves nearly the same performance as Linear. In summary, conducting model selection for the contrast functions yields better estimates of $\text{\slshape g}^{\text{opt}}$, and the prior preference for one particular candidate model can be reasonably incorporated into the model selection by combining $\widehat{R}_{cv}$ with its estimated variance.

\begin{table}[H]
\centering
\caption{\label{tab:DTR_accuracy} The average decision accuracy with the standard deviation shown in brackets and the percentage of the tree-based model being selected in the overall 200 replications under different methods.} \vspace{1mm}
\begin{tabular}{c c r r r r r}
  \hline \vspace{-4mm} \\ 
 &   & \multicolumn{3}{c}{accuracy} & \multicolumn{2}{c}{model selection [tree\%]}\\
\cmidrule(rl){3-5} \cmidrule(rl){6-7}
\multicolumn{1}{c}{case}&\multicolumn{1}{c}{method}  & \multicolumn{1}{c}{stage 1} & \multicolumn{1}{c}{stage 2}  & \multicolumn{1}{c}{both} & \multicolumn{1}{c}{stage 1} & \multicolumn{1}{c}{stage 2} \\ \hline \vspace{-2mm} \\ 
\multirow{4}{*}{i} &$\text{MCCV}_p$  & 0.9633 (0.0147)  & 0.8923 (0.0121)& 0.8579 (0.0185) & 0 &  4.5\\
&$\text{MCCV}$    & 0.9636 (0.0150) & 0.9058 (0.0251) & 0.8715 (0.0285)& 0 &  44.5\\
&Linear             & 0.9634 (0.0147) & 0.8911 (0.0107) &  0.8567 (0.0174)  &-  & -  \\
&Tree                & 0.8356 (0.0429)  &0.9212 (0.0305) & 0.7684 (0.0466)  &-  & - \\ \vspace{0.2mm} \\

\multirow{4}{*}{ii} &$\text{MCCV}_p$ &  0.8983 (0.0119)  & 0.9572 (0.0185) & 0.8582 (0.0206) & 1.5 &  0 \\
&$\text{MCCV}$  & 0.9027 (0.0202) & 0.9572 (0.0185) & 0.8625 (0.0242)  & 31.0 &  0 \\
&Linear    & 0.8979 (0.0111) & 0.9572 (0.0185) &  0.8578 (0.0199)  &-  & -  \\
&Tree     & 0.9084 (0.0322)  &0.8427 (0.0427) & 0.7638 (0.0511)    &-  & - \\ \vspace{0.2mm} \\

\multirow{4}{*}{iii} &$\text{MCCV}_p$  &  0.9624 (0.0164) & 0.9671(0.0126) & 0.9305 (0.0192) & 0 & 100 \\
&$\text{MCCV}$  &  0.9624 (0.0164) & 0.9671(0.0126) & 0.9305 (0.0192)  & 0 & 100  \\
&$\text{Linear}$  &  0.9392 (0.0287) & 0.8683(0.0035) &  0.8140 (0.0262)  & - & - \\
&$\text{Tree}$   & 0.8320 (0.0421)   & 0.9671(0.0126) &  0.8047 (0.0421)  & - & -  \\ \vspace{0.2mm} \\

\multirow{4}{*}{iv} &$\text{MCCV}_p$   &  0.9656 (0.0101) & 0.9199 (0.0460) &  0.8882 (0.0457) & 100 & 0 \\
&$\text{MCCV}$   &  0.9656 (0.0101) & 0.9199 (0.0460) &  0.8882 (0.0457) & 100 & 0 \\
&Linear     & 0.8730 (0.0042)  &0.9199  (0.0460)& 0.8020 (0.0424)  & -&- \\
&Tree   & 0.9642 (0.0110) & 0.7294 (0.1037)&  0.7021 (0.1023)     &- &- 
\\ \hline

\end{tabular}
\end{table}

\section{Real data analysis}
In this section, we apply the proposed model selection-assisted optimal DTR estimation to the data collected from the Sequenced Treatment Alternatives to Relieve Depression (STAR*D) study, which is a randomized clinical trial of outpatients with major depressive disorder \citep{rush2004sequenced}. The trial involved four levels/stages and proceeded as follows. All participants were assigned to citalopram at level 1. Those who did not achieve sufficient improvement, as assessed by the clinician based on the 16-item Quick Inventory of Depressive Symptomatology ($\text{QIDS-C}_{16}$, the higher the worse), then entered level 2 and were assigned to one of four switch options (switch from citalopram to sertraline, bupropion,
venlafaxine or cognitive therapy) or one of three augmentation options (augment citalopram with bupropion, buspirone or cognitive therapy) at the beginning of level 2. Those who received cognitive therapy (either switch or augment options) at level 2 without sufficient improvement then entered level 2A and were randomized to one of two switch options (switch to venlafaxine or bupropion) at the beginning of level 2A. Participants without sufficient improvement at level 2 or level 2 plus level 2A (if the cognitive therapy was assigned at level 2 and no sufficient improvement was achieved at level 2) entered level 3 and were assigned to one of two switch options (switch to mirtazapine or nortriptyline) or one of two augmentation options (augment with lithium or thyroid hormone) at the beginning of level 3. Those without sufficient improvement at level 3 entered level 4 and were randomized to one of two switch options (switch to ranylcypromine or to the combination of mirtazapine and venlafaxine). The treatment assignments between switch and augmentation strategies at level 2 and 3 were based on patients' preference while the randomized assignment was made among substrategies within switch or augmentation. At each treatment level, clinical visits were scheduled at weeks 0, 2, 4, 6, 9 and 12 during which symptom severity such as $\text{QIDS-C}_{16}$ was assessed. Participants with intolerance to current treatment or minimal reduction in $\text{QIDS-C}_{16}$ were encouraged to move to the next treatment level earlier.  Sufficient improvement is defined as $\text{QIDS-C}_{16}\leq 5$ for at least two weeks and at the same time without intolerable side effects. For those achieving partial (not sufficient) improvement at week 12 or exhibiting sufficient improvement only at week 12, two additional visits were scheduled to identify whether sufficient improvement could be achieved and sustained. These resulted in different duration times between participants at each level. Additionally, participants with sufficient improvement moved to naturalistic follow-up instead.

Following the analysis in \cite{schulte2014q}, we restrict the treatment action to switch and augmentation in level 2 (taking level 2A as a part of level 2) and level 3 irrespective of specific switch/augmentation options, i.e., $A_j=1$ for switch and $A_j=0$ for augmentation with $j=1,2$ indicating  stage 1 (level 2) and  stage 2 (level 3) respectively. We define the final outcome $Y$ to be the average of the negative QIDS-$\text{C}_{16}$ scores at the end of the available stages, i.e., $Y=(1-I_{\text{sim}})(-S_2-S_3)/2-I_{\text{sim}}S_2$, where $I_{\text{sim}}$ is the indicator of whether sufficient improvement was achieved at the end of stage 1 (level 2) and $S_j$ is the QIDS-$\text{C}_{16}$ score at the end of level $j$ (i.e. stage $j-1$) for $j=2,3$. We aim to estimate the optimal dynamic treatment regime in this setting which involves $K=2$ decision points for patients failing to achieve sufficient improvement with citalopram at level 1 in the STAR*D trial. Note that there were participants dropping out of the study when moving to the next level, i.e., who refused to enter the next level though no sufficient improvement was achieved at the current level. After deleting the dropouts, there are 815 participants in stage 1 (level 2): 329 of them moved to stage 2 (level 3), for
whom the observed records are $(L_1, A_1, L_2, A_2, Y)$. The others (486 participants with sufficient improvement achieved at the end of stage 1 (level 2)) entered
the naturalistic follow-up after level 2, for whom the observed records are $(L_1, A_1, Y)$. In addition to baseline covariates such as age, gender and $S_0$, the QIDS-$\text{C}_{16}$ score at the beginning of level 1, we include in $L_1$ the time spent in level 1, i.e. $\text{time}_1$, the QIDS-$\text{C}_{16}$ score at the end of level 1, i.e. $S_1$, and the slope of  QIDS-$\text{C}_{16}$ score over level 1, i.e. $S_{1s}=(S_1-S_0)/\text{time}_1$. As for $L_2$, we similarly include  $\text{time}_2$ the time spent in level 2 (stage 1), $S_2$, and $S_{2s}=(S_2-S_1)/\text{time}_2$ the slope of QIDS-$\text{C}_{16}$ score over level 2 (stage 1).

To investigate the treatment effect function $C_2(\cdot)$ in stage 2 (level 3), we in fact perform analysis on participants with $I_{\text{sim}}=0$ and use $-S_3$ as the response variable. The candidate models considered for $C_2(\cdot)$ and their performances in terms of $\widehat{R}_{cv}$ are reported in Table \ref{tab:stard_stage2}. Additional to the honest causal tree estimator which has been implemented throughout the paper, we also consider the adaptive tree estimator without pruning in this section to illustrate the performance of $\widehat{R}_{cv}$. Basically, the honest tree estimator differs from the adaptive tree estimator in i) the splitting criterion where the honest criterion penalizes small leaf size; and ii) the approach in estimating the magnitude of the treatment effect in each leaf where the honest approach splits the training set into two parts, one for constructing the tree and the other for estimating the treatment effects given the constructed tree structure \citep{athey2016recursive}. The numbers inside the brackets in Table \ref{tab:stard_stage2} are the estimated standard deviations of $\widehat{R}_{cv}$'s derived from the estimated correlation $\hat{\rho}^{\text{adj}}$.  

As can be seen from Table \ref{tab:stard_stage2}, the linear model with `gender' (and the intercept) performs best (lowest $\widehat{R}_{cv}$), although, broadly, the performance gaps (relative to the estimated standard deviations of $\widehat{R}_{cv}$) across candidate models are not wide (except for the `adaptive, no pruning' tree model). Further, we compare the best performing model and the constant model directly through setting  $U_i(D_{tr})=[\widetilde{Y}_i-\hat{\tau}_{gen}(H_{2i};D_{tr})]^2-[\widetilde{Y}_i-\hat{\tau}_{const}(H_{2i};D_{tr})]^2$ in $\widehat{R}_{cv}$ to investigate the degree of heterogeneity in $C_2(\cdot)$. Such comparison is motivated by the findings from \cite{yu2024contemporary} that most CATE estimators fail to outperform the constant estimator based on a large-scale benchmark study. Our result, as shown in the last row in Table \ref{tab:stard_stage2}, indicates that incorporating the heterogeneity by gender outperforms the constant estimator in terms of the value of $\widehat{R}_{cv}$, but there is insufficient evidence to show that this outperformance is of significance.
\begin{table}[H]
\centering
\caption{\label{tab:stard_stage2} Model selection for the contrast function in stage 2 in STAR*D} \vspace{1mm}
\begin{tabular}{c c r r }
\hline 
\multicolumn{2}{c}{candidate models}&\multicolumn{1}{c}{$\widehat{R}_{cv}$ (SD)}  & \multicolumn{1}{c}{$\hat{\rho}^{\text{adj}}$} \\ \hline \vspace{-2mm} \\ 
\multirow{5}{*}{linear} 
&  1            &47.27 (3.6808) &0.1051 \\ 
&$1+S_{2s}$       &47.60 (3.5695) &0.0973\\
&$1+S_{1s}+S_{2s}$&48.07 (3.6055) &0.0970 \\ 
&\textbf{1+gender}         &45.26 (3.2607) &0.0924 \\ 
&$\text{full}^{\ast}$  &47.39 (3.3962) &0.0887 \\ \vspace{1mm} \\
\multirow{2}{*}{tree}  
&adaptive, no pruning&56.13 (3.8706)& 0.1123\\
& honest, pruning    &48.80 (3.8669)&  0.1035\\ \vspace{1mm} \\
& 1+gender \textit{vs.} 1  & $-2.007$ (1.5076) &0.1655 \\
\hline 
\end{tabular}
\begin{tablenotes} \footnotesize
\item $^\ast$ full: $1+S_{1s}+S_{2s}+S_1+S_2+\text{gender}+\text{age}$
\end{tablenotes}
\end{table}

To proceed, we adopt the linear model `$1+\text{gender}$' in estimating the contrast function in stage 2. The estimated contrast function is given by $\widehat{C}_2(H_2)=-3.1337+2.9187\times I(\text{gender}=\text{male})$. That is, the treatment effect in stage 2 (level 3), conditional on the available history $H_2$, is negative for the whole population. Thus the treatment effect is quantitative and not qualitative, implying augmentation as the treatment recommendation for all participants in stage 2 (level 3), i.e, $\hat{\text{\slshape g}}^{\text{opt}}_2(H_2)=0$. However, the beneficial effect of augmentation in stage 2 is different between males and females: females benefit more from  augmentation than males. The value function is then updated as follows 
\begin{equation*}
\begin{aligned}
\widehat{V}_2&=(1-I_{\text{sim}})\left(-S_2+\left[-S_3+(\hat{\text{\slshape g}}^{\text{opt}}_2(H_2)-A_2)\widehat{C}_2(H_2)\right]\right)/2-I_{\text{sim}}S_2\\
&=(1-I_{\text{sim}})\left(-S_2+\left[-S_3-A_2\widehat{C}_2(H_2)\right]\right)/2-I_{\text{sim}}S_2,
\end{aligned}
\end{equation*}
which serves as the response variable in stage 1. 

To estimate the contrast function $C_1(\cdot)$  in stage 1, we consider the candidate estimators shown in Table \ref{tab:stard_stage1}. The linear model with `$\text{time}_1$' as the covariate performs best in terms of  $\widehat{R}_{cv}$ although the gain in performance relative to the constant estimator is not significant. Fitting the linear model `$1+\text{time}_1$' to all participants in stage 1 (level 2), we get $\widehat{C}_1(H_1)= 1.7634-0.0316 \times \text{time}_1$; that is, individuals with time spent in level 1 more than $ 1.7634/0.0316=55.8$ days (614 over 815 participants) should augment citalopram in stage 1 (level 2). Otherwise, switching is recommended. Here the treatment effect is qualitative.

In summary, our estimated optimal dynamic treatment regime for patients who cannot achieve sufficient improvement during level 1 with citalopram is to switch at the beginning of level 2 if they spend 55.8 days or less at level 1 (usually indicating a minimal reduction in the severity of symptoms or clear intolerance toward citalopram); otherwise, augment citalopram with additional treatment(s). Those entering level 3 without sufficient improvement during level 2 should augment their current treatment at the beginning of level 3. Moreover, as suggested by the last rows in Tables \ref{tab:stard_stage2} and \ref{tab:stard_stage1}, assigning all participants to augmentation at levels 2 and 3 may also be an effective strategy. Similar results have been reported in \cite{rush2007star} and \cite{menza2006star}.

\begin{table}[H]
\centering
\caption{\label{tab:stard_stage1} Model selection for the contrast function in stage 1 in STAR*D} \vspace{1mm}
\begin{tabular}{c c r r r r }
\hline 
\multicolumn{2}{c}{candidate models}&\multicolumn{1}{c}{$\widehat{R}_{cv}$ (SD)}  & \multicolumn{1}{c}{$\hat{\rho}^{\text{adj}}$} & \\ \hline \vspace{-2mm} \\ 
\multirow{5}{*}{linear} 
&  1                  &36.70 (2.3747) &0.1543 \\ 
&$1+S_{1s}$             &36.50 (2.3777) &0.1572 \\
&1+$S_{1s}+\text{time}_1$&36.45 (2.3550) & 0.1527\\ 
&$\textbf{1+time}_1$     &36.40 (2.3344) &0.1492\\ 
&$\text{full}^{\ast}$  &37.64 (2.4623) &0.1496\\ \vspace{1mm} \\
\multirow{2}{*}{tree}  
&adaptive, no pruning &48.08 (3.1351) &0.1644\\
&  honest, pruning   &37.08 (2.4934) &0.1550  \\ \vspace{1mm} \\
& $1+\text{time}_1$ \textit{vs.} 1  & $-0.303$ (0.3666) & 0.0838\\
\hline 
\end{tabular}
\begin{tablenotes}\footnotesize
\item $^\ast$ full: $1+S_{1s}+S_0+S_1+\text{time}_1+\text{gender}+\text{age}$
\end{tablenotes}
\end{table}

\section{Discussion}

In this paper, we firstly propose a data-adaptive  method to estimate the variance of $\widehat{R}_{cv}$, which allows us to quantify and account for the uncertainty in the model selection procedure to identify the best model/estimator for CATE under a given sample size.  The method is built on the fact that the risks $\widehat{R}_{1},\dots,\widehat{R}_{J}$ obtained from different splittings in MCCV are exchangeable with a common correlation $\rho$. By tackling the estimation of $\rho$, we derive an estimator of the variance of  $\widehat{R}_{cv}$, which, as demonstrated in the numerical studies in Section \ref{sec:sim_var}, is close to the Monte Carlo variance and outperforms the existing variance estimators. Secondly, by applying the proposed approach to the contrast functions in DTRs, we develop the model selection-assisted A-learning for estimating optimal DTRs as outlined in Algorithm \ref{algorithm}. The model selection at the $k$th decision point can be performed through hypothesis testing by combining $\widehat{R}_{cv,k}$ with its estimated variance. Incorporating model selection on contrast functions in A-learning helps to mitigate against model misspecification and leads to $\hat{\text{\slshape g}}^{\text{opt}}$ with improved decision accuracy and  value $\mathbb{E}[Y^{\ast}(\hat{\text{\slshape g}}^{\text{opt}})]$.

Rather than selecting the best candidate, SuperLearner aims to find the best combination of candidate estimators based on a performance metric \citep{van2007super,montoya2023optimal}. When using the negative of the expected regime-specific outcome $-\mathbb{E}[Y^{\ast}(\hat{\text{\slshape g}})]$ as the metric, the library of candidate estimators in SuperLeaner can include both the indirect estimators (e.g., estimators for Q-functions or contrast functions) and the direct estimators (e.g., OWLs) for optimal DTRs, whereas the candidate estimators in our method are restricted to those for contrast functions. For the estimation of $\mathbb{E}[Y^{\ast}(\hat{\text{\slshape g}})]$, the cross-validated targeted maximum likelihood estimator (CV-TMLE) has been developed \citep{van2015targeted}. As an ensemble approach, SuperLearner exhibits attractive performance in prediction by integrating multiple individual learners. However, the interpretation of its result is not straightforward.

The model selection procedure proposed in this paper also applies to the high-dimensional setting where variable selection accompanied estimators, for example, penalized A-learning in \cite{shi2018high}, should be considered as candidate estimators. However, constructing the pseudo-label/surrogate $\widetilde{Y}$ by pairing the most similar individuals based on Euclidean (or Mahalanobis) distance of the high-dimensional $L$ will lead to a noisy $\widehat{R}_{cv}$, which will be reflected in the value of $\text{Var}[\widehat{R}_{cv}]$. Sufficient dimension reduction has been considered in the supplement of \cite{rolling2014model}. In fact, for a treated individual $i\in I_{T}$, the missing outcome is $Y_i^{\ast}(0)$ and thus we aim to identify the individual $i^{\prime}\in I_{C}$ whose observed outcome $Y_{i^{\prime}}=Y^{\ast}_{i^{\prime}}(0)$ is close to $Y_i^{\ast}(0)$. That is, the matching for  $i\in I_{T}$ should be based on the control group mean function $\mu_0(L)=\mathbb{E}[Y|L,A=0]$ \citep{gao2021assessment}. Similarly, matching for control individuals should be based on the treatment group mean function $\mu_1(L)=\mathbb{E}[Y|L,A=1]$. In addition to the matching distance, \cite{gao2021assessment} also considered the matching structure to ensure that as many individuals as possible are used and that no individuals are overused in matching. It is expected that constructing $\widetilde{Y}$ based on elaborate design of matching distances and matching structures will improve the performance of  $\widehat{R}_{cv}$ in model assessment and model selection. Moreover, other constructions of $\widetilde{Y}$ than through matching, as discussed in \cite{schuler2018comparison}, can  be utilized and our proposed variance estimator and the model selection framework can be straightforwardly applied.

\label{Bibliography}
\bibliographystyle{apalike} 

\bibliography{Bibliography} 

\section*{Appendix A: Proof of Proposition 1}
As $\rho=\text{Cov}(\widehat{R}_1,\widehat{R}_2)/\text{Var}(\widehat{R}_1)$, we derive expressions for $\text{Var}(\widehat{R}_1)$ and $\text{Cov}(\widehat{R}_1, \widehat{R}_2)$ in the following. Following the technique employed in \cite{NIPS1999}, we introduce the indicator function $\mathbb{I}_j(i)$  which equals to 1 if $D_i$ falls in the validation set under the $j$th splitting, i.e., $\mathbb{I}_j(i)=1$ if $i \in I_{val}^{(j)}$; and $\mathbb{I}_j(i)=0$ otherwise. Then for $\text{Var}(\widehat{R}_1)$, we have  
$$\text{Var}(\widehat{R}_1)=\mathbb{E}\left\{\text{Var}[\widehat{R}_1\mid\mathbb{I}_1(\cdot)]\right\}+\text{Var}\left\{\mathbb{E}[\widehat{R}_1\mid \mathbb{I}_1(\cdot)]\right\}.\eqno(\text{A}.1)$$
For the first term in (\text{A}.1), we have 
$$
\begin{aligned}
& \operatorname{Var}\left[\widehat{R}_1 \mid \mathbb{I}_1(\cdot)\right] \\
=&\operatorname{Var}\left[\frac{1}{n_2} \sum_{i=1}^n \mathbb{I}_1(i) U_i(D_{tr}^{(1)}) \mid \mathbb{I}_1(\cdot)\right] \\
=&\frac{1}{n_2^2} \Bigg\{\sum_{i=1}^n \mathbb{I}_1(i) \operatorname{Var}\left[U_i(D_{tr}^{(1)}) \mid \mathbb{I}_1(\cdot)\right] +2\sum_{i <i^{\prime}} \mathbb{I}_1(i) \mathbb{I}_1(i^{\prime}) \operatorname{Cov}\left(U_i\left(D_{tr}^{(1)}\right), U_{i^{\prime}}\left(D_{tr}^{(1)}\right)\mid \mathbb{I}_1(\cdot) \right)\Bigg\}\\
=&\frac{1}{n_2^2}\Bigg\{\operatorname{Var}\left[U_i(D_{tr}^{(1)})\right]\sum_{i=1}^n \mathbb{I}_1(i)+ \operatorname{Cov}\left(U_i\left(D_{tr}^{(1)}\right), U_{i^{\prime}}\left(D_{tr}^{(1)}\right)\right)2\sum_{i <i^{\prime}} \mathbb{I}_1(i) \mathbb{I}_1(i^{\prime}) \Bigg\}\\
=&\frac{1}{n_2}\operatorname{Var}\left[U_i(D_{tr}^{(1)})\right]+\frac{2}{n_2^2}\operatorname{Cov}\left(U_i\left(D_{tr}^{(1)}\right), U_{i^{\prime}}\left(D_{tr}^{(1)}\right)\right)\sum_{i <i^{\prime}} \mathbb{I}_1(i) \mathbb{I}_1(i^{\prime}). 
\end{aligned}
$$
Then, 
\begin{align*}
&\mathbb{E}\left\{\text{Var}[\widehat{R}_1\mid\mathbb{I}_1(\cdot)]\right\}\\
=&\frac{1}{n_2}\operatorname{Var}\left[U_i(D_{tr}^{(1)})\right]+\frac{2}{n_2^2}\operatorname{Cov}\left(U_i\left(D_{tr}^{(1)}\right), U_{i^{\prime}}\left(D_{tr}^{(1)}\right)\right)\sum_{i <i^{\prime}} \mathbb{E}\left[\mathbb{I}_1(i) \mathbb{I}_1(i^{\prime})\right]\\
=& \frac{1}{n_2}\operatorname{Var}\left[U_i(D_{tr}^{(1)})\right]+\frac{2}{n_2^2}\operatorname{Cov}\left(U_i\left(D_{tr}^{(1)}\right), U_{i^{\prime}}\left(D_{tr}^{(1)}\right)\right)\binom{n}{2}\frac{1 \times \binom{n-2}{n_2-2}}{\binom{n}{n_2}}\\
=&\frac{1}{n_2}\operatorname{Var}\left[U_i(D_{tr}^{(1)})\right]+\frac{n_2-1}{n_2}\operatorname{Cov}\left(U_i\left(D_{tr}^{(1)}\right), U_{i^{\prime}}\left(D_{tr}^{(1)}\right)\right),\tag{A.2}
\end{align*}
where the second equality holds because the probability of both $i$ and $i^{\prime}$ falling in the validation set in one splitting is $\binom{n-2}{n_2-2}/\binom{n}{n_2}$.
As for the second term in (\text{A}.1), note that 
$$\mathbb{E}\left[\widehat{R}_1 \mid \mathbb{I}_1(\cdot)\right]=\mathbb{E}\left[\frac{1}{n_2} \sum_{i=1}^n \mathbb{I}_1(i) U_i(D_{tr}^{(1)}) \mid \mathbb{I}_1(\cdot)\right]=\mathbb{E}[U_i(D_{tr}^{(1)})],$$
which does not depend on $\mathbb{I}_1(\cdot)$. Therefore, we have $\text{Var}\left\{\mathbb{E}[\widehat{R}_1\mid \mathbb{I}_1(\cdot)]\right\}=0$, and $\text{Var}(\widehat{R}_1)$ has the expression given by (A.2). 

We now deal with $\text{Cov}(\widehat{R}_1,\widehat{R}_2)$.
Applying the conditional covariance formula to $\text{Cov}(\widehat{R}_1,\widehat{R}_2)$ gives
\begin{align*}
& \operatorname{Cov}\left(\widehat{R}_1, \widehat{R}_2\right)\\
= & \mathbb{E}\left[\operatorname{Cov}\left(\widehat{R}_1, \widehat{R}_2 \mid I_1(\cdot), I_2(\cdot)\right)\right]+\operatorname{cov}\left(\mathbb{E}\left[\widehat{R}_1 \mid I_1(\cdot), I_2(\cdot)\right], \mathbb{E}\left[\widehat{R}_2\mid I_1(\cdot), I_2(\cdot)\right]\right).\tag{A.3}
\end{align*}
For the conditional variance in the first term in (A.3), we have
\begin{align*}
&\operatorname{Cov}\left(\widehat{R}_1, \widehat{R}_2 \mid I_1(\cdot), I_2(\cdot)\right) \\
=&\operatorname{Cov}\left(\frac{1}{n_2} \sum_{i=1}^n I_1(i) U_i\left(D_{tr}^{(1)}\right), \frac{1}{n_2} \sum_{i=1}^{n} I_2(i) U_i\left(D_{tr}^{(2)}\right) \mid I_1(\cdot), I_2(\cdot)\right) \\
=&\frac{1}{n_2^2} \sum_{i, j} \operatorname{Cov}\left(I_1(i) U_i\left(D_{tr}^{(1)}\right), I_2(j) U_j\left(D_{tr}^{(2)}\right) \mid I_1(\cdot), I_2(\cdot)\right) \\
=&\frac{1}{n_2^2}\left[\sum_{i=1}^n I_1(i) I_2(i) \operatorname{Cov}\left(U_i\left(D_{tr}^{(1)}\right), U_i\left(D_{tr}^{(2)}\right)\right)+\sum_{i \neq j}I_1(i) I_2(j)\operatorname{Cov}\left(U_i\left(D_{tr}^{(1)}\right), U_i\left(D_{tr}^{(2)}\right)\right)\right].
\end{align*}
It follows that
\begin{align*}
&\mathbb{E}\left[\operatorname{Cov}\left(\widehat{R}_1, \widehat{R}_2 \mid I_1(\cdot), I_2(\cdot)\right)\right]\\
=&\frac{1}{n_2^2} \sum_{i=1}^n \frac{n_2}{n} \frac{n_2}{n} \operatorname{Cov}\left(U_i\left(D_{tr}^{(1)}\right), U_i\left(D_{tr}^{(2)}\right)\right)+\frac{1}{n_2^2} \sum_{i\neq j} \frac{n_2}{n} \frac{n_2}{n} \operatorname{Cov}\left(U_i\left(D_{tr}^{(1)}\right), U_j\left(D_{tr}^{(2)}\right)\right)\\
=&\frac{1}{n}\operatorname{Cov}\left(U_i\left(D_{tr}^{(1)}\right), U_i\left(D_{tr}^{(2)}\right)\right)+\frac{2}{n_2^2} \binom{n}{2}\frac{n_2^2}{n^2}\operatorname{Cov}\left(U_i\left(D_{tr}^{(1)}\right), U_j\left(D_{tr}^{(2)}\right)\right)\\
=&\frac{1}{n}\operatorname{Cov}\left(U_i\left(D_{tr}^{(1)}\right), U_i\left(D_{tr}^{(2)}\right)\right)+\frac{n-1}{n} \operatorname{Cov}\left(U_i\left(D_{tr}^{(1)}\right), U_j\left(D_{tr}^{(2)}\right)\right),
\end{align*}
where the first equality holds because the splittings in MCCV are independent of each other and that for each individual, the probability of falling in the validation set is $n_2/n$. Moreover, the second term in (A.3) equals zero as $\mathbb{E}[\widehat{R}_j \mid I_1(\cdot), I_2(\cdot)]=\mathbb{E}[\widehat{R}_j]$ for $j=1,2$. Finally, we arrive at the following result for $\rho$:
$$\rho=\frac{\frac{1}{n}\operatorname{Cov}\left(U_i\left(D_{tr}^{(1)}\right), U_i\left(D_{tr}^{(2)}\right)\right)+\frac{n-1}{n} \operatorname{Cov}\left(U_i\left(D_{tr}^{(1)}\right), U_j\left(D_{tr}^{(2)}\right)\right)}{\frac{1}{n_2}\operatorname{Var}\left[U_i(D_{tr}^{(1)})\right]+\frac{n_2-1}{n_2}\operatorname{Cov}\left(U_i\left(D_{tr}^{(1)}\right), U_{i^{\prime}}\left(D_{tr}^{(1)}\right)\right)}=\frac{\frac{1}{n}\rho_1+\frac{n-1}{n}\rho_3}
{\frac{1}{n_2}+\frac{n_2-1}{n_2}\rho_2},$$
where $\rho_1=\text{corr}(U_i(D_{tr}^{(1)}),U_i(D_{tr}^{(2)}))$ with $i \in I_{val}^{(1)}\cap I_{val}^{(2)}$, i.e., the correlation between $U$'s evaluated with the same individual under different training sets/splittings; $\rho_2=\text{corr}(U_i(D_{tr}^{(1)}),U_{i^{\prime}}(D_{tr}^{(1)}))$ with $i, i^{\prime}\in I_{val}^{(1)}$ and $i\neq i^{\prime}$, i.e., the correlation between $U$'s evaluated with different individuals under the same splitting; and $\rho_3=\text{corr}(U_i(D_{tr}^{(1)}),U_{i^{\prime}}(D_{tr}^{(2)}))$ with $i \in I_{val}^{(1)}$, $i^{\prime} \in I_{val}^{(2)}$ and $i\neq i^{\prime}$, i.e., the correlation between $U$'s evaluated at different individuals under different splittings.

\section*{Appendix B: simulation under a regression setting}
\setcounter{table}{0}
\renewcommand{\thetable}{B\arabic{table}}
Data are generated as follows:
$Y_i=m(X_i)+\epsilon_i=2+2X_{1i}+\epsilon_i$ with $X_{1i}\overset{iid}{\sim} \text{Ber}(0.5)$ and $\epsilon_i \overset{iid}{\sim} N(0,5^2)$, for $i=1,\dots,200$. Additionally, we generate a nuisance covariate $X_{2i} \overset{iid}{\sim} N(5,2^2)$ for $i=1,\dots,200$ and $X_i=(X_{1i},X_{2i})^{\top}$. Moreover, $X_1$, $X_2$ and $\epsilon$ are independent of each other. We assess the performance of the four different models/estimators shown in Table \ref{tab:B1} by setting $U_i(D_{tr})=[Y_i-\hat{m}(X_i;D_{tr})]^2$ for a given estimator $\hat{m}$ when calculating $\widehat{R}_{cv}$; and also compare the performance between two estimators, e.g., $\hat{m}_1$ and $\hat{m}_2$, by setting  $U_i(D_{tr})=[Y_i-\hat{m}_{1}(X_i;D_{tr})]^2-[Y_i-\hat{m}_{2}(X_i;D_{tr})]^2$. Apart from the value of $\widehat{R}_{cv}$, we are interested in the variance of $\widehat{R}_{cv}$ and the performance of $\text{var}_{\rho=0.2}$ which estimates $\text{Var}(\widehat{R}_{cv})$ by approximating $\rho$ with the splitting ratio $q$ (here $q=0.2$) in MCCV. We denote the Monte Carlo variance of $\widehat{R}_{cv}$ by $\text{var}^{\ast}$ in Table \ref{tab:B1}. As shown in Table \ref{tab:B1}, $\text{var}_{\rho=0.2}$ may underestimate or overestimate  $\text{Var}(\widehat{R}_{cv})$ and the bias is substantial in some cases.
\begin{table}[H]
\centering
\caption{\label{tab:B1} Results based on 200 Monte Carlo repetitions.} \vspace{1mm}
\begin{tabular}{c c r r r r}
\hline \vspace{-3mm} \\
\multicolumn{2}{c}{models}&\multicolumn{1}{c}{$\widehat{R}_{cv}$}  & \multicolumn{1}{c}{$\text{var}^{\ast}$} & \multicolumn{1}{c}{$\text{var}_{\rho=0.2}$} \\ \hline \vspace{-2mm} \\ 
\multirow{3}{*}{linear} 
&  1               &26.36 & 8.270 & 6.870  \\ 
&$1+X_1$            & 25.49 & 7.290 &6.535\\
&$1+X_1+X_2$      & 25.66 & 7.558  &6.627\\ \vspace{1mm} \\
\multirow{1}{*}{tree}  
&adaptive, no pruning &29.62 &10.21 & 9.183\\ \vspace{1mm} \\
&$1+X_1$ \textit{vs.} $1+X_1+X_2$ & $-0.170$ & 0.0382 &0.108\\
&$1+X_1$ \textit{vs.} tree        & $-4.174$ & 0.994 & 3.320\\
\hline 
\end{tabular}
\end{table}

\end{document}